\documentclass[entropy,article,submit,pdftex,moreauthors]{Definitions/mdpi} 
\renewcommand{\linenumbers}{}
\firstpage{1} 
\pubvolume{1}
\issuenum{1}
\articlenumber{0}
\pubyear{2023}
\copyrightyear{2022}
\datereceived{} 
\daterevised{}
\dateaccepted{} 
\datepublished{} 
\hreflink{https://doi.org/} 

\usepackage[utf8]{inputenc} 
\usepackage{times}			
\usepackage[T1]{fontenc}

\graphicspath{{figs/}}

\usepackage{amssymb,amsmath,amsthm,graphicx,microtype}
\usepackage{array}
\usepackage{mathrsfs}
\usepackage{stmaryrd}

\usepackage{algpseudocode}

\newcommand{\cRM}[1]{\MakeUppercase{\romannumeral #1}}	
\newcommand{\crm}[1]{\romannumeral #1}

\usepackage{tikz}
\usepackage{circuitikz}

\usepackage{scalerel}
\usepackage{pict2e}
\usepackage{tkz-euclide}
\usetikzlibrary{calc}
\usetikzlibrary{patterns,arrows.meta}
\usetikzlibrary{shadows}
\usetikzlibrary{external}

\usepackage{pgfplots}
\pgfplotsset{compat=newest}
\usepgfplotslibrary{statistics}
\usepgfplotslibrary{fillbetween}

\usepackage{caption,subcaption} 

\usepackage{xcolor}

\definecolor{red}{rgb}{1.,0.,0.}
\definecolor{green}{rgb}{0.,0.6,0.}
\definecolor{blue}{rgb}{0.,0.,0.6}

\newcommand{\Id}{\mathrm{Id}}

\renewcommand{\P}{\ensuremath\mathbb{P}}
\renewcommand{\epsilon}{\varepsilon}

\Title{An information theoretic condition for perfect reconstruction}

\TitleCitation{An information theoretic condition for perfect reconstruction}

\Author{Idris Delsol$^{1}$, Olivier Rioul\orcidA{} $^{1}$, Julien Béguinot\orcidB{} $^{1}$, Victor Rabiet$^{1,2}$, and Antoine Souloumiac$^{3}$}

\AuthorNames{Idris Delsol, Olivier Rioul, Julien Béguinot, Victor Rabiet, and Antoine Souloumiac}

\AuthorCitation{Delsol, I., Rioul, O., Béguinot, J., Rabiet, V., and Souloumiac, A.}

\address{%
$^{1}$ \quad 
LTCI, Télécom Paris, Institut Polytechnique de Paris, 91120 Palaiseau, France; 
firstname.name@telecom-paris.fr\\
$^{2}$ \quad 
DMA, ENS, 75005 Paris, France;
victor.rabiet@telecom-paris.fr, victor.rabiet@ens.fr\\
$^{3}$ \quad 
Paris-Saclay University, CEA, LIST, 91120, Palaiseau, France;
antoine.souloumiac@cea.fr
}

\corres{Correspondence: olivier.rioul@telecom-paris.fr} 

\abstract{%
A new information theoretic condition is presented for reconstructing a discrete random variable~$X$ based on the knowledge of a set of discrete functions of~$X$.
The reconstruction condition is derived from Shannon's 1953 lattice theory with two entropic metrics of Shannon and Rajski.
Because such a theoretical material is relatively unknown and appears quite dispersed in different references, we first provide a synthetic description (with complete proofs) of its concepts, such as total, common and complementary informations. Definitions and properties of the two entropic metrics  are also fully detailed and shown compatible with the lattice structure.
A new geometric interpretation of such a lattice structure is then investigated that leads to a necessary (and sometimes sufficient) condition for reconstructing the discrete random variable $X$ given a set $\{ X_1,\ldots,X_{n} \}$ of elements in the lattice generated by $X$.
Finally, this condition is illustrated in five specific examples of perfect reconstruction problems:
reconstruction of a symmetric random variable from the knowledge of its sign and absolute value,  reconstruction of a word from a set of linear combinations,  reconstruction of an integer from its prime signature (fundamental theorem of arithmetic) and from its remainders modulo a set of coprime integers (Chinese remainder theorem), and reconstruction of the sorting permutation of a list from a minimal set of pairwise comparisons. 
}


\begin{document}

\begin{quotation}\centering
 \noindent A movement is accomplished in six stages\\
 And the seventh brings return.\\
 The seven is the number of the young light\\
 It forms when darkness is increased by one.\\\smallskip
 Change returns success\\
 Going and coming without error.\\
 Action brings good fortune.\\
 Sunset, sunrise.
\\\hfill\sl Syd Barrett, Chapter 24 (Pink Floyd). 
\end{quotation}

\section{Introduction}

We consider the problem of perfectly reconstructing a discrete random variable~$X$, based on the knowledge of a finite set $X_{1}$, $X_{2}$,..., $X_{n}$ of deterministic processings or transformations of~$X$, denoted $f_i$ such that $X_i = f_i (X)$.
Intuitively, the components $X_i$ are assumed to carry only a partial amount of the ``information'' present in $X$ and perfect reconstruction of $X$ would only be possible if the combination of the ``informations'' in $X_{1}$, $X_{2}$,..., $X_{n}$ is enough to contain all the original ``information'' in $X$.
Such intuitive considerations expressed in the language of information is very common in signal processing and in many other scientific fields; but they were never mathematically formalized  as far as the authors know.
This article aims at formalizing precisely this trivial and vague intuition. Such a task implies, in particular, an accurate definition of ``information''.

The classical Shannon's 1948 information theory~\cite{Shannon48} cannot really answer this question as it is rather a theory of the measure of information rather than of the information itself.
Fortunately, a ``true information'' theory has also been developed by Claude Shannon in a relatively unknown 1953 article~\cite{Shannon53}, that is \emph{not} what is generally referred to as ``Shannon's information theory.''
Said briefly, the information is there defined as an equivalence class of discrete random variables. A partial order on a set of classes allows one to build a lattice structure called the \emph{information lattice}, which is made metric by the introduction of two related entropic distances.



``\textsl{Claude [Shannon] did not like the term `\textit{information theory}'}\/'' recalls Robert Fano, a colleague of Shannon's working at MIT, who died almost century-old just seven years ago. In one of his last interviews~\cite{Fano01}, he says: ``\textsl{You see, the term 'information theory' suggests that it's a theory about information, but it's not. It's about the transmission of information, not the information. Many people just didn't understand that.}''
Fano is of course referring to Shannon's famous theory in his 1948 seminal paper~\cite{Shannon48} which he entitled ``\,a mathematical theory of  \emph{communication}\,'' -- not information. 
But very early on, it is the term ``information'' that prevails. The entropy  $H(X)$ of a discrete random variable  $X$ is presented as the measure of  ``\emph{information} contained in  $X$'', and the notion of \emph{mutual information} $I(X;Y)$ between two variables $X$ and $Y$, introduced precisely by the same Robert Fano in his course at MIT~\cite{Fano52}, quickly became central to the teaching of the theory.
Moreover, the very first historical article on the theory, barely three years after its birth (!) is entitled ``\textit{A history of the theory of information}''~\cite{Cherry51}.

This sudden craze for ``information'' in the early 1950s eventually became somewhat of a bore for Shannon, who in 1956, in his famous editorial \textit{The Bandwagon}~\cite{Shannon56} warned against the excesses of such popularity: ``\textsl{It will be all too easy for our somewhat artificial prosperity to collapse overnight when we realize that the use of a few exciting words like information, entropy, redundancy, does not solve all our problems. }''

Under these conditions, it is understandable that Shannon wanted to go further: If several, unrelated, random variables can have the same \emph{quantity} of information $H$, how can information itself be defined? Shannon presented a very brief summary of his findings (without proofs) at the international congress of mathematicians (ICM) in 1950~\cite{Shannon50} and in a small, relatively unknown article~\cite{Shannon53} published in 1953 in the very first issue of what was to become the {IEEE Transactions on information theory}.




The remainder of this article is organized as follows.
Section~\ref{information} presents in detail the Shannon theory of the lattice of information with complete proofs,  
and Section~\ref{sec-distances} does the same for the two entropic distances proposed respectively by Shannon and Rajski.
The corresponding geometric point of view is further developed in Section~\ref{geometry}.
Two conditions of perfect reconstruction, a necessary one and a sufficient one, are then derived in Section~\ref{reconstruction}.
Finally, the condition is applied to five specific examples in Section~\ref{application}.

Sections~\ref{information}, \ref{sec-distances} and Subsection~\ref{alignment} are a deepening of the article \cite{Gretsi22} previously published (in French) by four of the authors. 

%
%
%

\section{What is Information? A Detailed Study of Shannon's Information Lattice }\label{information}

For simplicity, we consider with Shannon discrete random variables $X$ which take a \emph{finite} number of values in some alphabet $\mathcal{X}$. 
This amounts to considering all the random variables $X: \Omega\to \mathcal{X}$ defined on a given probability space $(\Omega,\mathcal{P}(\Omega),\P)$ where the underlying universe $\Omega$ is finite and $\mathcal{P}(\Omega)$ is the power set of $\Omega$.


\subsection{Definition of the ``True'' Information}

Quite arguably, the \emph{information} contained in a discrete source or random variable $X$ should not be confused with the ``measure of quantity of information'' such as the entropy $H(X)$.
Shannon's idea~\cite{Shannon53} is that this information contained in $X$ should in fact be defined as $X$ itself! Of course, any reversible encoding of $X$ must be regarded as the \emph{same} information, since one moves from one representation to another without loss of information. This amounts, in modern language, to the following definition:
\begin{Definition}[``True'' Information]\label{def:info}
The \emph{information} (contained in) $X$ is the equivalence class of $X$ for the equivalence relation:
\begin{equation}
X\equiv Y \iff Y=f(X) \text{ and } X=g(Y) \text{ a.s. (almost surely)} 
\end{equation}
 for two deterministic functions $f$ and $g$.
\end{Definition}

\begin{proof}

Relation $\equiv$ is evidently reflexive (take $f$, $g$ be the identity function) and symmetric (by permuting the roles of $f$ and $g$ in the definition). It is also transitive by composition: If $X \equiv Y$ and $Y \equiv Z$, there exists $f$, $g$, $h$ and $k$ such that $Y = f(X)$, $X = g(Y)$ and $Y = h(Z)$, $Z = k(Y)$ a.s.; then $X = g(h(Z)) = g \circ h(Z)$ and $Z = k \circ f(X)$ a.s.
%
\end{proof}



\begin{Proposition}\label{prop-one}
$X \equiv Y$ iff (if and only if) there exists a bijective function $h$ such that $Y = h(X)$ a.s.
\end{Proposition}

\begin{proof}
If $X \equiv Y$, then there exist two deterministic functions $f$ and $g$ such that, $X = f(Y)$ and $Y = g(X)$ a.s. Thus, $X = f(g(X))$ a.s. Then, for every value $X=x$ with non-zero probability, $f \circ g(x) = x$. Hence, $f\circ g$ coincides with the identity function a.s. Since the problem is symmetric in $X$ and $Y$, $g \circ f$ also coincides with the identity function a.s. Thus, $h=g$ is bijective from the set of values that $X$ can take with non-zero probability to the set of values that $Y$ can take with non-zero probability, and we have $Y = g(X)= h(X)$ a.s.

Conversely, if $Y = h(X)$ a.s. with bijective $h$, then $X = h^{-1}(Y)$ a.s., hence $X \equiv Y$.
\end{proof}

As suggested by Rajski~\cite{Rajski61}, the equivalence between $X$ and $Y$ can be characterized by way of their joint probability matrix:

\begin{Proposition}[Matrix Characterization]
If we restrain $\Omega$ to the elements of non-zero probability measure, $X \equiv Y$ iff the matrix of joint probabilities $\mathbb{P}(X=x, Y=y)$ is a permutation matrix.
\end{Proposition}

\begin{proof}
By Proposition~\ref{prop-one}, $X \equiv Y$ iff there exists a bijective function $h$ such that $Y = h(X)$ a.s. 
Thus to each outcome of $X$ corresponds exactly one outcome of $Y$ and \textit{vice versa}, which is equivalent to saying that the matrix of joint probabilities is a permutation matrix.
\end{proof}

In the following, we shall note (without possible confusion) $X$ the equivalence class of the variable $X$, and thus $X=Y$ the equality between the two classes $X$ and $Y$ (rather than $X\equiv Y$).

With this definition, it is clear that the equivalence relation is compatible with any functional relation $Y=f(X)$. If $f$ is not bijective, it is tempting to say that there is \emph{less} information in $Y$ than in $X$. Hence the following partial order.
\begin{Definition}[Partial Order]\label{def:ordre}
\begin{equation}
X\geq Y \iff Y=f(X)\vspace*{-.5ex} \text{ a.s.} 
\end{equation}
for some deterministic function $f$.
\end{Definition}
We also write $Y\leq X$. We are not necessarily considering real valued variables, so the order $X\geq Y$ has nothing to do with the order in $\mathbb{R}$.

\begin{Proposition}
The relation $\geq$ is indeed a partial order on the set of equivalence classes of the relation $\equiv$ defined above.
\end{Proposition}

\begin{proof}
We first show that the relation $\equiv$ is compatible with the relation $\geq$. 
Let $X_{1}$, $X_{2}$ and $Y_{1}$, $Y_{2}$ be such that $X_{1} \equiv X_{2}$ and $Y_{1} \equiv Y_{2}$. Then if $X_{1} \geq Y_{1}$, there exists a deterministic function $f$ such that $Y_{1} = f(X_{1})$ a.s.. Since $X_{1} \equiv X_{2}$, there exists a bijective $h$ such that $X_{1} = h(X_{2})$ a.s., hence $Y_{1} = f \circ h(X_{2})$ a.s. and $X_{2} \geq Y_{1}$. Likewise, since $Y_{1} \equiv Y_{2}$, there exists a bijective $g$ such that $Y_{2} = g(Y_{1})$ a.s., so $Y_{2} = g \circ f \circ h(X_{2})$ a.s., hence $X_{2} \geq Y_{2}$.
This shows that the relation $\geq$ is well defined on the set of equivalence classes of the relation $\equiv$.

We now show that $\geq$ is indeed a partial order:
\begin{itemize}
    \item \emph{Reflexivity}: $X = \Id(X)$ so $X \geq X$.
    \item\emph{Antisymmetry}: If $X \geq Y$ and $Y \geq X$, $X = f(Y)$ a.s. and $Y = g(X)$ a.s. for deterministic functions $f$ and $g$, so $X \equiv Y$.
    \item \emph{Transitivity}: If $X \geq Y$ and $Y \geq Z$, then there exist two deterministic functions $f$ and $g$ such that: $Z = g(Y)$ a.s. and $Y = f(X)$ a.s. Then $Z = g(f(X))$ a.s., hence $X \geq Z$.\qedhere
\end{itemize}
\end{proof}

\subsection{Structure of the Information Lattice: Joint Information; Common Information}

Beyond the partial order, Shannon~\cite{Shannon53} established the natural mathematical structure of information: It is a \emph{lattice}, i.e. two variables $X,Y$ always admit a maximum $X\lor Y$ and a minimum $X\land Y$. Let us recall that these quantities (necessarily unique if they exist) are defined by the relations
\begin{equation}
\begin{split}
(X\leq Z \text{ and }  Y\leq Z) \iff X\lor Y \leq Z,  \\
(X\geq Z \text{ and }  Y\geq Z) \iff X\land Y \geq Z.
\end{split}
\end{equation}
Shannon, in his paper~\cite{Shannon53}, used Boolean notations instead, $X+Y$ for $X\lor Y$ and $X\!\cdot\!Y$ for $X\land Y$.

\begin{Proposition}[Joint Information]
The joint information $X\lor Y$ of $X$ and $Y$ is the random pair $X\lor Y=(X,Y)$.
\end{Proposition}
\begin{proof}
If $X$ and $Y$ are functions of $Z$, then the pair $(X,Y)$ is also a function of $Z$. Conversely, since $X$ and $Y$ are functions of $(X,Y)$, if $(X,Y)$ is a function of $Z$ then so are $X$ and $Y$. 
\end{proof}

The definition of $X\wedge Y$ (\textit{common information}) is more difficult and was not made explicit by Shannon. Following G\'acs and K\"orner~\cite{GacsKorner73}, let us adopt the following definition:
\begin{Definition}
We say that $x\in\mathcal{X}$ and $y\in\mathcal{Y}$ \emph{communicate}, denoted by $x\sim y$, if there exists a path $x y_1 x_1 y_2 \cdots y_nx_n y$ in which all transitions are of non zero probability: $\P(X=x,Y=y_1)>0$, $\P(Y=y_1, X=x_1)>0$, \ldots, \hbox{$\P(X=x_n, Y=y)>0$}.
\end{Definition}

\begin{Proposition}
The relation $\sim$ is an equivalence relation on the set of pairs $(x,y)$ for which $\mathbb{P}(X = x) > 0$ and $\mathbb{P}(Y = y) > 0$. 
\end{Proposition}

\begin{proof} \emph{Reflexivity} is obvious.
\begin{itemize}
\item \emph{Symmetry}: If $x \sim y$, taking the path  $x\ldots y$ in the other direction and we have $y \sim x$. 
\item \emph{Transitivity}: If $x_{1} \sim y_{1}$, $y_{1} \sim x_{2}$ and $x_{2} \sim y_{2}$, then there exists a path from $x_{1}$ to $y_{1}$, another from $y_{1}$ to $x_{2}$ and a last one from $x_{2}$ to $y_{2}$, whose transitions are of non-zero probabilities. The concatenated path from $x_{1}$ to $y_{2}$ has non zero transition probabilities, hence $x_{1} \sim y_{2}$. \qedhere
\end{itemize}
\end{proof}

\begin{Definition}[Communication Class]
The communication class $C(x,y)$ is the equivalence class of $(x,y)$ where $x\sim y$.
\end{Definition}

\begin{Proposition}[Common Information]
The common information $X\land Y$ of $X$ and $Y$ is $X\land Y=C(X,Y)$.
\end{Proposition}
\begin{proof}
If $Z=f(X)=g(Y)$ a.s. then $Z$ is constant for each pair $(x,y)$ such that $x\sim y$; in other words $Z$ is a function of the class $C(X,Y)$.
\end{proof}

\begin{Remark}
In order to compute the common information between $X$ and $Y$ in practice, one has to fully determine communication classes, which is only possible if there is a finite number of classes, each of which containing a finite number of elements. In other words, $X$ and $Y$ should take a finite number of values. This is the reason why we restrict ourselves to  finitely valued variables in this paper.
\end{Remark}

\begin{Remark}
As in any lattice, $X\leq Y$ is equivalent to saying that $X\lor Y=Y$ or that $X\land Y=X$.
\end{Remark}

\subsection{Computing Common Information}
As shown in the previous section, the definition of common information is not a simple one, but one can compute it efficiently using the following algorithm. Given two variables $X$ and $Y$, this algorithm turns the joint probability matrix of $(X,Y)$ into a \emph{block-diagonal} matrix where each block corresponds to each communication class.

Let $X$ and $Y$ be two random variables taking values in $\mathcal{X}$ and $\mathcal{Y}$, respectively. Consider the graph $G = (V,E)$ whose vertices $V$ are $\mathcal{X} \cup \mathcal{Y}$, and such that vertices $x$ and $y$ of $V$ are connected by an edge if and only if $\mathbb{P}(X=x, Y=y) > 0$. Hence, $G$ is fully described by the joint probability matrix $\mathbb{P}_{X,Y}$. Furthermore, this is a bipartite graph (no edge connects two vertices $x_{1}$, $x_{2}$ belonging to $\mathcal{X}$ or two vertices $y_{1}$, $y_{2}$ belonging to $\mathcal{Y}$). 

Then, the communication classes $C(X,Y)$ correspond to the \emph{connected components} of~$G$. Indeed, a connected component~$C$ is a subset of~$V$ such that each of its elements is accessible to all the others by a path in the subgraph $(C,E)$. So for any two vertices $x$, $y$ in the connected component~$C$, there exists $y_{1}$, $x_{1}$, \ldots, $y_{k}$, $x_{k}$ such that all the edges $(x,y_{1})$, $(y_{1},x_{1})$,\ldots, $(y_{k},x_{k})$, $(x_{k},y)$ belong to~$E$, that is all the transition probabilities between these vertices are non-zero, which is equivalent to saying that they belong to the same communication class.
Now, it is known that the connected components of~$G$ can be determined by a depth-first search. 

We propose an algorithm, whose pseudo-code is given in Fig.~\ref{alg_common_info}, that takes as input the joint probability matrix $\mathbb{P}_{X,Y}$ and outputs a bloc-diagonal form of $\mathbb{P}_{X,Y}$ representing the common information $X \wedge Y$, an array storing the permutation of the columns of $\mathbb{P}_{X,Y}$ and an array storing the permutation of the rows $\mathbb{P}_{X,Y}$. Since the matrix $\mathbb{P}_{X,Y}$ is sufficient to fully describe $G$, we adapt the depth-first search algorithm to browse the rows and columns of the matrix $\mathbb{P}_{X,Y}$ to find which of its rows and columns must be swapped in order to write this matrix into a block-diagonal form. In this algorithm (Fig.~\ref{alg_common_info}), the $i$th row of $\mathbb{P}_{X,Y}$ will be represented by the pair $(r,i)$ and the $j$th column by the pair $(c,j)$.

\begin{figure}
\begin{algorithmic}[1]
\State input $\mathbb{P}_{X,Y}$: $n_R\times n_C$ matrix   \Comment{Joint probability matrix}
\State $\sigma_R\gets$  array of integers of length $n_R$  \Comment{Rows permutation vector} 
\State $\sigma_C\gets$ array of integers of length $n_C$ \Comment{Columns permutation vector} 
\State $S\gets$ empty stack \Comment{Stack contains row indices $(r,i)$ or column indices $(c,j)$}
\State push $(r,0)$ into stack $S$  \Comment{First row put into stack}
\State $\text{\it bottom}\gets 1$ \Comment{Bottommost row index not yet assigned}
\State $\text{\it up}\gets n_R-1$ \Comment{Upmost row index that may have nonzero entries}
\State $\text{\it left}\gets 0$ \Comment{Leftmost column index not yet assigned}
\State $\text{\it right} \gets n_C-1$ \Comment{Rightmost column index that may have nonzero entries}
\While{There is an unmarked row or column}
    \While{$S$ is not empty}
        \State $(s,i) \gets S.pop()$ \Comment{The $pop()$ operation removes the top stack element and returns it.}
        \If{$(s,i)$ is not marked}
            \State mark $(s,i)$ 
            \If{$s = r$} \Comment{Current index $i$ is a row index}
                \For{$\text{\it left}\leq j \leq \text{\it right}$}  \Comment{Scan all columns}
                    \If{$\mathbb{P}_{X,Y}(i,j) > 0$}
                        \State push $(c,j)$ into stack $S$
                        \State $\sigma_C[j] \gets \text{\it left}$; swap columns $\text{\it left}$ and $j$ in $\mathbb{P}_{X,Y}$
                        \State $\text{\it left} \gets \text{\it left}+1$
                    \EndIf
                \EndFor
                \If{all entries on $i$th row are zeros}
                    \State $\sigma_R[i] \gets \text{\it up}$; swap rows $i$ and $\text{\it up}$ in $\mathbb{P}_{X,Y}$
                    \State $\text{\it up} \gets \text{\it up} - 1$
                \EndIf
            \Else \Comment{Current index $i$ is a column index}
                \For{$\text{\it bottom} \leq j \leq \text{\it up}$} \Comment{Scan all rows}
                    \If{$\mathbb{P}_{X,Y}(j,i) > 0$}
                        \State push $(r,j)$ into stack $S$
                        \State $\sigma_R[j] \gets \text{\it bottom}$; swap rows $\text{\it bottom}$ and $j$ in $\mathbb{P}_{X,Y}$
                        \State $\text{\it bottom} \gets \text{\it bottom}+1$
                    \EndIf
                \EndFor
                \If{all entries on $i$th column are zeros}
                    \State $\sigma_C[i] \gets \text{\it right}$; swap columns $i$ and $\text{\it right}$ in $\mathbb{P}_{X,Y}$
                    \State $\text{\it right} \gets \text{\it right} - 1$
                \EndIf
            \EndIf
        \EndIf
    \EndWhile \Comment{Empty Stack}
    \If{there is an unmarked $i$th row $(r,i)$}
        \State push $(r,i)$ into stack $S$
    \ElsIf{there is an unmarked $j$th column $(c,j)$}
        \State push $(c,j)$ into stack $S$
    \EndIf
\EndWhile \Comment{All rows and columns marked}
\State \Return $\mathbb{P}_{X,Y}$, $\sigma_R$, $\sigma_C$
\end{algorithmic}
\caption{Algorithm to compute the common information}\label{alg_common_info}
\end{figure}

The complexity of this algorithm can be determined as follows. Let $n = \text{Card}(\mathcal{X}) + \text{Card}(\mathcal{Y})$ be the sum of the alphabet sizes on which $X$ and $Y$ take their values, i.e. the sum of the number of rows of $\mathbb{P}_{X,Y}$ and the number of columns of $\mathbb{P}_{X,Y}$. 
The algorithm passes through each row and column at most once. Indeed, for the index of a row or column to enter the stack, it must be \emph{unmarked}, but as soon as we put it on the stack, we mark it. Then, each time the index of a row or a column is unstacked, we look at each coefficient of the corresponding row or column. 
Therefore, our algorithm looks at each coefficient of the joint probability matrix $\mathbb{P}_{X,Y}$ exactly once.
Four elementary operations are performed each time we cross a nonzero coefficient. Thus, the algorithm complexity is quadratic in $n$. 

Notice that the output of our algorithm gives a visualization of the common information: The stochastic matrix $\P(X=x,Y=y)$ is written, after permutation of rows/columns, in the "block diagonal" form 
\begin{equation}\label{eq-matrix}
\P_{X,Y} = 
\begin{pmatrix}
C_1   \\
        & C_2 &            &  {\text{\Large \bf 0}}\\
        &         & \ddots \\
{\text{\Large \bf 0}}  &         &           & C_k  \\
    &   &   &   & 0 \\
    &   &   &   &   & \ddots \\
    & & & & &  & 0\\
    
\end{pmatrix} 
\end{equation}
where $k$, the number of blocks, is maximal. The $k$ rectangular matrices then represent the $k$ different equivalence classes, the probability $\P(C(X,Y)=i)$ being the sum of all entries in block $C_i$.

\subsection{Boundedness and Complementedness: Null, Total, and Complementary Information}

\begin{Proposition}[Null Information; Total Information]
The information lattice is \emph{bounded}, i.e., it admits a minimum $0$ and maximum $1$, such that for any $X$, $0\leq X\leq 1$.
\begin{itemize}
    \item The minimal element $0$ (``null information'') is the equivalence class of all \emph{deterministic} variables. Thus $X=0$ means that $X$ is a deterministic variable.
    \item The maximal element $1$ (``total information'') of the lattice is the equivalence class of the identity function $\Id$ on $\Omega$.
\end{itemize}
\end{Proposition}
\begin{proof}
If $X$ is any random variable and $Z=c$ a.s. is any deterministic variable, then it is clear that $Z = f(X)$ where $f$ is the constant function $c$. Letting $0$ be the equivalence class of constant variables, one has $0\leq X$ for all $X$.

Also, for any random variable $X$, $X = X \circ \Id$, hence $X \leq \Id$. Letting $1$ be the equivalence class of the identity function on $\Omega$, one has $X\leq 1$ for all $X$. 
\end{proof}

\begin{figure}[htb!]
    \centering
  \vspace*{-.15cm}
   \includegraphics[width=\textwidth]{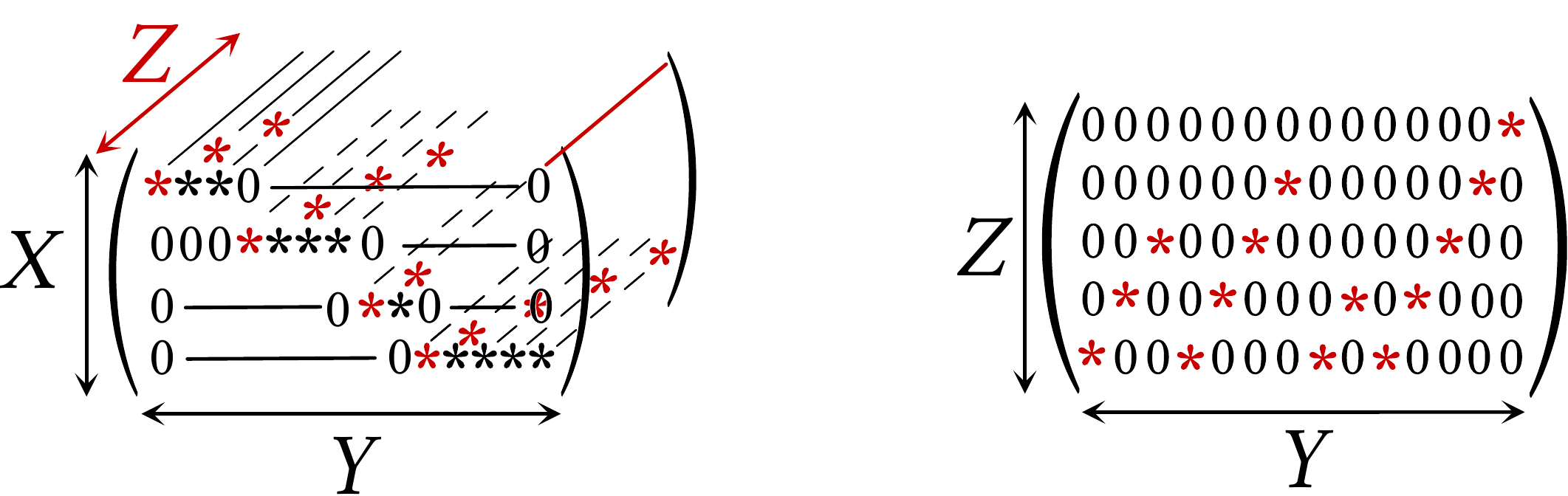}
      \vspace*{-.3cm}
    \caption{\small Construction of the complementary information $Z$ allowing to pass from $X$ to $Y$. The stochastic tensor of $(X,Y,Z)$ representing $\P_{X,Y,Z}$ has nonzero entries marked in red. The distribution $\P_Z$ of $Z$ is obtained by marginalizing the tensor on the $Z$ axis. }
    \label{fig-tenseur}
\end{figure}

\begin{Proposition}[Complementary Information]\label{prop-complemented}
The information lattice is \emph{complemented}, i.e., any $X\leq Y$ admits a  complement $Z$ (``complementary information'') such that $X\lor Z=Y$ and $X\land Z=0$.
\end{Proposition}
\noindent This $Z$ is the information missing from $X$ to obtain $Y$: It allows $Y$ to be reconstructed from $X$ without requiring more information than necessary. Shannon in~\cite{Shannon53} did not say how to determine it. The following proof gives an explicit construction:
\begin{proof}
Since $X\leq  Y$, we simply have $X=X\land Y = C(X,Y)$. Thus, a given class $C(X,Y)=x$ has only one value $X=x$ per class, corresponding in general to several values of $Y$, say, $y^x_1,y^x_2, \ldots,y^x_{k_x}$. Now let $Z\in\{1,\ldots,k_X\}$ be the unique index such that $Y=Y^X_Z$. 

By construction, $Z\leq X\lor Y=Y$, and since $X\leq Y$, one also has $X\lor Z\leq Y$. But the formula $Y=Y^X_Z$ shows that $Y\leq X\lor Z$, hence equality $X\lor Z=Y$ holds.

Finally, the value $Z=1$ connects each pair $(x,z)$, so there is only one class according to $(X,Z)$, i.e. $X\land Z=0$.
\end{proof}
This construction can be visualized on the stochastic tensor of $(X,Y,Z)$ described in Fig.~\ref{fig-tenseur}.

\begin{Remark}
The complementary information $Z$ is \emph{not} uniquely determined by $X$ and $Y$.
In the above construction, it depends on how the values of $Y$ are indexed by the class $X=x$.
\end{Remark}

\subsection{Computing the Complementary Information}
Given $X \leq Y$, the algorithm of Fig.~\ref{alg_complementary_info} determines a random variable $Z$ corresponding to the complementary information from $X$ to $Y$.
%
This algorithm takes as input the joint probability matrix $\mathbb{P}_{X,Y}$ in its bloc-diagonal form and outputs the tensor of the joint probability $\mathbb{P}_{X,Y,Z}$ where $X\lor Z=Y$ and $X\land Z=0$.
The tensor is built by spreading the non-zero coefficients of the joint probability matrix $\mathbb{P}_{X,Y}$ on the $Z$-axis as shown in Fig.~\ref{fig-tenseur}.

\begin{figure}[ht!]
\begin{algorithmic}[1]
    \State input $\mathbb{P}_{X,Y}$: $n_R\times n_C$ matrix \Comment{Joint probability matrix}
    \State $k \gets 0$ \Comment{Z index}
    \For{$0 \leq i < nR$}
        \For{$0 \leq j < nC$}
            \If{$\mathbb{P}_{X,Y}(i,j) > 0$}
                \State $\mathbb{P}_{X,Y,Z}(i,j,k) \gets \mathbb{P}_{X,Y}(i,j)$
                \State $k \gets k+1$
            \EndIf
        \EndFor
        \State $k \gets 0$
    \EndFor
\State \Return $\mathbb{P}_{X,Y,Z}$
\end{algorithmic}
\caption{Algorithm for computing the complementary information} \label{alg_complementary_info}
\end{figure}

The algorithm looks at each coefficient of the joint probability matrix $\mathbb{P}_{X,Y}$ exactly once and performs at most two elementary operations for each coefficient it processes. Therefore, it is quadratic in $n = \text{Card}(\mathcal{X}) + \text{Card}(\mathcal{Y})$ (since the number of coefficients in the matrix $\mathbb{P}_{X,Y}$ is quadratic in $n$). 

\subsection{Is the Information Lattice a Boolean Algebra?}

Interestingly, it was Shannon who, as early as 1938 in his master's thesis, used the \emph{Boolean algebra} to study relay-based circuits -- ``the most important master's thesis of the century'' for which Shannon received the Alfred Noble prize (not to be confused with the Alfred Nobel Prize!) in 1940. But alas, as Shannon noted, his information lattice is \emph{not} a Boolean algebra. It would have been one if it were \emph{distributive} ($\land$ distributive with respect to $\lor$ or vice versa), since a Boolean algebra is, by definition, a distributive complemented bounded lattice.  However:

\begin{Proposition}
The information lattice is not distributive.
\end{Proposition}

\begin{proof}[Indirect Proof]
In any Boolean algebra, the complement is unique. As seen above, this is not the case for the information lattice.
\end{proof}

\begin{proof}[Direct Proof]
As a direct second proof, we provide an explicit counterexample to distributivity. Consider the probability space $(\Omega,\mathcal{P}(\Omega),\P)$, where $\Omega = \{0, 1, 2, 3\}$ and $\P$ is the uniform probability measure, and define
$X(\omega)=0$ if $\omega$ is even, $X(\omega)=1$ otherwise.
%
%
%
%
Now let $Z_{1}$, $Z_{2}$ be given as in Table~\ref{tab:1} below.
%
%
%
%
As we read in the table, $(X\wedge Z_{1}) \vee (X \wedge Z_{2})=0$ is constant while $X \wedge (Z_{1} \vee Z_{2})$ is not. Therefore, $(X\wedge Z_{1}) \vee (X \wedge Z_{2}) \not = X \wedge (Z_{1} \vee Z_{2})$, and  the information lattice is not distributive.
\begin{table}[h!]
    \centering
        \caption{Computation of $X\wedge (Z_{1} \vee Z_{2})$ and of $(X\wedge Z_{1}) \vee (X \wedge Z_{2})$.}
    \label{tab:1}
        \begin{tabular}{|c|c|c|c|c|}
        \hline
        $\omega$ & 0 & 1 & 2 & 3 \\
        \hline
        $X$ & 0 & 1 & 0 & 1\\
        \hline
        $Z_{1}$ & 1 & 1 & 2 & 2\\
        \hline
        $Z_{2}$ & 2 & 1 & 1 & 2\\
        \hline
        $Z_{1}\vee Z_{2}$ & (1,2) & (1,1) & (2,1) & (2,2)\\
        \hline
        $X \wedge (Z_{1}\vee Z_{2})$ & 0 & 1 & 0 & 1\\
        \hline
        $X \wedge Z_{1}$ & 0 & 0 & 0 & 0\\
        \hline
        $X \wedge Z_{2}$ & 0 & 0 & 0 & 0\\
        \hline
        $(X\wedge Z_{1}) \vee (X \wedge Z_{2})$ & (0,0) & (0,0) & (0,0) & (0,0)\\
        \hline
        \end{tabular}
\end{table}
\end{proof}

\section{Metric Properties of the Information Lattice} \label{sec-distances}


\subsection{Information and Information Measures}

First of all, it is immediate to check the compatibility of the information lattice with respect to the entropy or the mutual information as measures of information. 


\begin{Proposition}
Entropy, conditional entropy and mutual information are compatible with the definition of information as an equivalence class.
\end{Proposition}

\begin{proof}\mbox{}
\begin{itemize}
\item \emph{Entropy}: If $X \equiv Y$, there exist functions $f$ and $g$ such that $Y = f(X)$ a.s., hence $H(Y) \leq H(X)$, and $X = g(Y)$ a.s., hence $H(X) \leq H(Y)$. Thus, $H(X) = H(Y)$.
\item \emph{Conditional entropy}: Let $X_{1} \equiv X_{2}$ with $f$ and $g$ two functions such that $X_{1} = f(X_{2})$ and $X_{2} = g(X_{1})$ a.s. Then $H(X_{1}|Y) = H(f(X_{2})|Y) \leq H(X_{2}|Y)$. Similarly, $H(X_{2}|Y) = H(g(X_{1})|Y) \leq H(X_{1}|Y)$. Therefore $H(X_{1}|Y) = H(X_{2}|Y)$. Finally, if $Y_{1} \equiv Y_{2}$ with  two functions $h$ and $k$ such that $Y_{1} = h(Y_{2})$ and $Y_{2} = k(Y_{1})$ a.s., then, $H(X|Y_{1}) = H(X|h(Y_{2})) \geq H(X|Y_2,h(Y_{2}))=H(X|Y_{2})$ and likewise $H(X|Y_{2}) = H(X|k(Y_{1})) \geq H(X|Y_1,k(Y_{1}))=H(X|Y_{1})$. Therefore $H(X|Y_{1}) = H(X|Y_{2})$.
\item \emph{Mutual information}: Since $I(X;Y) = H(X) - H(X|Y)$, compatibility follows from the two previous cases. \qedhere
\end{itemize}
\end{proof}

We then have some obvious connections:
\begin{Proposition}[Partial Order and Conditional Entropy]\label{prop-H}
\begin{equation}
X\leq Y \iff H(X|Y)=0 
\end{equation}
In particular, $H$ is ``order-preserving'' (greater information implies higher entropy):
\begin{equation}
X\leq Y \implies H(X)\leq H(Y). 
\end{equation}
Also, $X\leq Y$ with $H(X) = H(Y)$ implies $X = Y$.

Finally $H(X)\geq 0$ for all $X$, with equality $H(X)=0$ iff $X=0$.
\end{Proposition}
\begin{proof}
$H(X|Y)=0$ means that $H(X|Y=y)=0$ for all $y\in\mathcal{Y}$, which amounts to saying that $X$ is deterministic equal to $f(y)$ given $Y=y$. In other words $X=f(Y)$ a.s. 
We then have $H(X)=H(X)-H(X|Y)=I(X;Y)=H(Y)-H(Y|X)\leq H(Y)$. 

Next, suppose that $X\leq Y$ and $H(X) = H(Y)$.
By the chain rule, $H(X,Y)= H(X) + H(Y|X)= H(Y) + H(X|Y)$.
Therefore, it follows from the equality $H(X) = H(Y)$ that $H(Y|X) = H(X|Y)$. But since $X\leq Y$, $H(X|Y) = 0$, hence $H(Y|X) = 0$ also, that is, $Y \leq X$. This shows equivalence $Y = X$.

Finally, since $X\geq 0$ for all $X$, $H(X)\geq H(0)=0$ and it is well known that the entropy $H(X)$ is zero if and only if the variable $X$ is deterministic, that is, $X=0$.
\end{proof}


\subsection{Common Information vs. Mutual Information}

\begin{Proposition}
The entropy of the joint information is the joint entropy, i.e. $H(X \vee Y) = H(X,Y)$.
\end{Proposition}

\begin{proof}
Obvious since $X \vee Y = (X,Y)$.
\end{proof}

One may wonder by analogy with the usual Venn diagram in information theory (Fig.~\ref{fig-Venn}) if the entropy of joint information is equal to the mutual information: is it true that $H(X\land Y)=I(X;Y)$?
The answer is \emph{no}, as shown next. Proposition~\ref{prop-common} is implicit in~\cite{GacsKorner73}, and made explicit by Wyner in~\cite{Wyner75} who credits a private communication from Kaplan.

\begin{Proposition}\label{prop-common}
$H(X\land Y)\leq I(X;Y)$ always, with equality $H(X\land Y)= I(X;Y)$ iff one can write $X=(U,W)$ and $Y=(V,W)$ where $U$ and $V$ are conditionally independent given~$W$. 
\end{Proposition}

\begin{proof}
Let $W=X\land Y$. Since $W\leq X$ and $W\leq Y$, by complementarity we can write $X=W\lor U=(U,W)$ and $Y=W\lor V=(V,W)$. By the chain rule for mutual information, $I(X;Y)=I(U,W;V,W)=I(W;V,W)+I(U;V,W|W)=H(W)+I(U;V|W)\geq H(W)$ with equality iff $U$ and $V$ are conditionally independent given~$W$.
\end{proof}

\begin{Remark}
In particular, if $X$ and $Y$ are independent, they have a null common information $X\land Y=0$. However,  common information $H(X\land Y)$ can be far less~\cite{GacsKorner73} than mutual information $I(X;Y)$.
\end{Remark}

\begin{Remark}
Notice that the case of equality corresponds to the case where the matrix blocks $C_i$ in~\eqref{eq-matrix} are stochastic matrices of two independent variables $X,Y$ knowing $W=i$, i.e. matrices of rank $1$.
\end{Remark}

\begin{Remark}
Shannon's notion of common information should not be confused with the well known Wyner's acception of ``common information'', which is defined as the maximum of $I(X,Y;W)$ when $X$ and $Y$ are conditionally independent knowing $W$. This quantity is not less but \emph{greater} than the mutual information $I(X;Y)$~\cite{Wyner75}. 
\end{Remark}

\subsection{Submodularity of Entropy on the Information Lattice}

From the results in~\cite{Nakamura70} we can show that entropy is \emph{submodular} on the information lattice:

\begin{Proposition}[Submodularity of entropy]
$H(X \vee Y) + H(X \wedge Y) \leq H(X) + H(Y)$.
\end{Proposition}

\begin{proof}
Since $X\wedge Y\leq Y$, $H(Y)=H(Y, X\wedge Y) =H(X\wedge Y)+H(Y|X\wedge Y)$.
But since $X\wedge Y\leq X$, $H(Y|X\wedge Y)\geq H(Y|X\wedge Y,X)=H(Y|X)=H(X \vee Y) -H(X)$.
Combining gives the announced inequality.
\end{proof}

\begin{Remark}
Submodularity is in fact equivalent to the inequality $H(X \wedge Y)\leq I(X;Y)$ of Proposition~\ref{prop-common}, since  $H(X) + H(Y) - H(X \vee Y) =H(X) + H(Y) - H(X,Y)=I(X;Y)$.
\end{Remark}

\begin{Remark}
The submodularity property of entropy that is generally studied in the information theory literature is with respect to the set lattice (or algebra), where the entropy is that of a collection of random variables indexed by some index set (thus considered as a set function).
Such considerations have been greatly developed in recent years, see, e.g.,~\cite{Yeung08}. 
By contrast, it is the information lattice that is considered here. 
It can be easily shown using Proposition~\ref{prop-common} that the two notions of submodularity coincide for collections of independent random variables.
\end{Remark}

\subsection{Two Entropic Metrics: Shannon Distance; Rajski Distance}

Since $X=Y \iff (X\leq Y \text{ and } X\geq Y)$, according to Prop.~\ref{prop-H}, it suffices that $H(X|Y)+H(Y|X)=0$ in order for $X$, $Y$ to be equivalent: $X=Y$. Shannon~\cite{Shannon53} noted that this defines a \emph{distance} which makes the information lattice a \emph{metric} space:

\begin{Proposition}[Shannon's Entropic Distance]\label{prop-D}
$D(X,Y)=H(X|Y)+H(Y|X)$ is a distance over the information lattice.
\end{Proposition}
\begin{proof}
\mbox{}
\begin{itemize}
\item \emph{Positivity}: As just noted above, $D(X,Y) \geq 0$ vanishes only when $X = Y$.
\item \emph{Symmetry}: $D(X,Y) = D(Y,X)$ is obvious by commutativity of addition.
\item \emph{Triangular inequality}: First note that $H(X|Z) \leq H(X,Y|Z) = H(X|Y,Z) + H(Y|Z) \leq H(X|Y) + H(Y|Z)$. By permuting $X$ and $Z$, we also obtain that $H(Z|X) \leq H(Z|Y) + H(Y|X)$. Summing up the two inequalities, we obtain the triangular inequality $D(X,Z)=H(X|Z) + H(Z|X) \leq H(X|Y) + H(Y|X) + H(Y|Z) + H(Z|Y)=D(X,Y)+D(Y,Z)$. 
\qedhere
\end{itemize}
\end{proof}
It is interesting to note that this is not the only distance (nor the only topology). By normalizing $D(X,Y)$ by the joint entropy $H(X,Y)$, we obtain another distance metric:
\begin{Proposition}[Rajski's Entropic Distance~\cite{Rajski61}]\label{distance-Rajski}
$d(X,Y)=\dfrac{D(X,Y)}{H(X,Y)}$ (with the convention $d(0,0)=0$) is a distance taking values in $[0,1]$.
\end{Proposition}

Notice that normalization by $H(X,Y)$ is valid when $X$ and $Y$ are non-deterministic since $X \not = 0$ and $Y \not = 0$ implies $H(X,Y) > 0$. 

\begin{proof} 
First of all, symmetry $d(X,Y)=d(Y,X)$ is obvious and positivity follows from that of $D$. 
We follow Horibe~\cite{Horibe73} to prove the triangular inequality. One may always assume non deterministic random variables.
Observe that:
\begin{equation}\label{eq-rajski-denom-1}
\frac{H(X|Y)}{H(X,Y)} =\frac{H(X|Y)}{H(X|Y) + H(Y)} \geq \frac{H(X|Y)}{H(X|Y) + H(Y,Z)}=\frac{H(X|Y)}{H(X|Y) + H(Y|Z) + H(Z)}
\end{equation}
and
\begin{equation}\label{eq-rajski-denom-2}
 \frac{H(Y|Z)}{H(Y,Z)}=\frac{H(Y|Z)}{H(Y|Z) + H(Z)} \geq \frac{H(Y|Z)}{H(X|Y) + H(Y|Z) + H(Z)}.
\end{equation}
Summing \eqref{eq-rajski-denom-1} and  \eqref{eq-rajski-denom-2} 
yields
\begin{equation}
\frac{H(X|Y)}{H(X,Y)} + \frac{H(Y|Z)}{H(Y,Z)} \geq \frac{H(X|Y) + H(Y|Z)}{H(X|Y) + H(Y|Z) + H(Z)}. 
\end{equation}
Now, from the above proof of the triangular inequality of $D$, one has $H(X|Y) + H(Y|Z) \geq H(X|Z)$. Noting  that  $a \geq b > 0$ and $c \geq 0$ imply $\frac{a}{a+c} \geq \frac{b}{b+c}$, we obtain
\begin{equation}
\frac{H(X|Y) + H(Y|Z)}{H(X|Y) + H(Y|Z) + H(Z)} 
\geq \frac{H(X|Z)}{H(X|Z) + H(Z)} \label{demo-ineq-triangulaire-2} 
= \frac{H(X|Z)}{H(X,Z)}. 
\end{equation}
Therefore,
\begin{equation} \label{eq:a}
    \frac{H(X|Y)}{H(X,Y)} + \frac{H(Y|Z)}{H(Y,Z)} \geq \frac{H(X|Z)}{H(X,Z)}.
\end{equation}
Permuting the roles of $X$ and $Z$ gives
\begin{equation} \label{eq:b}
    \frac{H(Y|X)}{H(X,Y)} + \frac{H(Z|Y)}{H(Y,Z)} \geq \frac{H(Z|X)}{H(X,Z)}.
\end{equation}
Summing \eqref{eq:a} and \eqref{eq:b}, we conclude that $d(X, Y) + d(Y, Z) \geq d(X,Z)$.
\end{proof}

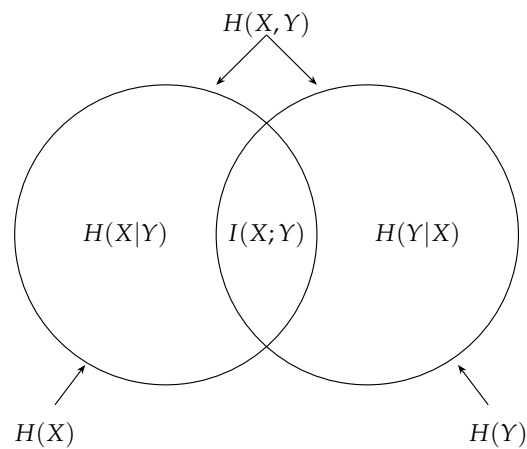
\begin{figure}[!ht]
\centering
    \resizebox{0.5\textwidth}{!}{%
    \begin{circuitikz}
    \tikzstyle{every node}=[font=\LARGE]
    \draw [, line width=0.8pt ] (5,11.25) circle (3.75cm);
    \draw [, line width=0.8pt ] (10,11.25) circle (3.75cm);
    \node [font=\LARGE] at (2,6.25) {$H(X)$};
    \node [font=\LARGE] at (13.25,6.25) {$H(Y)$};
    \node [font=\LARGE] at (7.5,16.5) {$H(X,Y)$};
    \node [font=\LARGE] at (7.5,11.25) {$I(X;Y)$};
    \node [font=\LARGE] at (4,11.25) {$H(X|Y)$};
    \node [font=\LARGE] at (11.25,11.25) {$H(Y|X)$};
    \draw [ line width=0.8pt, -Stealth] (2.25,7) -- (3,8);
    \draw [ line width=0.8pt, -Stealth] (13,7) -- (12.25,8);
    \draw [ line width=0.8pt, -Stealth] (7.5,16.25) -- (8.75,15);
    \draw [ line width=0.8pt, -Stealth] (7.5,16.25) -- (6.25,15);
    \end{circuitikz}
    }%
    \caption{Usual Venn diagram in information theory.}\label{fig-Venn}
\end{figure}

\begin{Remark}
Rajski's distance between two variables $X$ and $Y$ can be visualized as the \emph{Jaccard distance} between the region corresponding to $X$ and the region corresponding to $Y$ in the Venn diagram of Fig.~\ref{fig-Venn}. 
The Jaccard (or Jaccard-Tanimoto) distance~\cite{Jaccard01} between two sets $A$ and $B$ is defined by $d_{J}(A,B) = \frac{|A \Delta B|}{|A \cup B|}$, where $\Delta$ is the symmetric difference between $A$ and $B$. Thus, if $A$ and $B$ are respectively the regions corresponding to $X$ and to $Y$ in the Venn diagram, we have: $H(X,Y) = |A \cup B|$, $H(X|Y) = |A\backslash B|$ and $H(Y|X) = |B\backslash A|$. Thus $\frac{H(X|Y) + H(Y|X)}{H(X,Y)} = \frac{|(A\backslash B)\cup(B\backslash A)|}{A\cup B} = \frac{|A \Delta B|}{|A \cup B|}$.
\end{Remark}

\subsection{Dependency Coefficient}
From the Rajski distance, we can define a quantity which measures the \emph{dependence} between two non-deterministic (i.e., nonzero)  random variables $X$ and $Y$. 

\begin{Definition}[Dependency Coefficient]\label{def-dependency}
For all non-zero elements $X$,~$Y$ of the information lattice, their \emph{dependency coefficient} is 
$\rho(X,Y) = 1 - d(X,Y) \in [0,1]$.
\end{Definition}

\begin{Proposition} The dependency coefficient can be seen as a normalized mutual information:
$\rho(X,Y) = \dfrac{I(X;Y)}{H(X,Y)}$.
\end{Proposition}

\begin{proof}
One has $1 - d(X,Y) =  1 - \dfrac{H(X|Y) + H(Y|X)}{H(X,Y)}=\dfrac{H(X,Y) - H(X|Y) - H(Y|X)}{H(X,Y)}$ where the numerator $=H(X) + H(Y|X) - H(X|Y) - H(Y|X)=I(X;Y)$.
\end{proof}

\begin{Proposition}
For non-deterministic $X$ and $Y$, one has $0 \leq \rho(X,Y) \leq 1$ where $\rho(X,Y)=0$ vanishes (equivalently, $d(X,Y) = 1$)  iff $X$ and $Y$ are independent
and $\rho(X,Y)=1$ (equivalently, $d(X,Y) = 0$) iff $X=Y$ are equivalent.
    \label{property-rho}
\end{Proposition}

\begin{proof}
If $X$ and $Y$ are independent, then $I(X;Y) = 0$, hence $\rho(X,Y) = 0$.
If $X$ and $Y$ are equivalent, then $d(X,Y) = 0$ and $\rho(X,Y) = 1 - d(X,Y) = 1$.
Since $0 \leq I(X;Y) \leq H(X) \leq H(X,Y)$, $0 \leq \rho(X,Y) = \frac{I(X;Y)}{H(X,Y)} \leq 1$.
\end{proof}

\begin{Remark}
The property of $\rho$ in proposition \ref{property-rho} is similar to the usual property of the linear correlation coefficient. However, while two independent random variables have zero correlation (but not conversely),  the corresponding converse property holds for the dependence coefficient since 
two  random variables are independent if and only if $\rho(X,Y) = 0$.
\end{Remark}

\subsection{Discontinuity and Continuity Properties}

Perhaps the biggest flaw in Shannon's lattice information theory~\cite{Shannon53} is that the different constructions of elements in the lattice (e.g., common and complementary information) do not actually depend on the \emph{values} of the probabilities involved, but only on whether they are equal to or different from zero.
Thus, a small perturbation on probabilities can greatly influence the results. As an illustration we have the following
\begin{Proposition}[Discontinuity of common information]
The application $(X,Y)\mapsto X\land Y$ is \emph{discontinuous} in the metric lattice with distance $D$ (or $d$).
\end{Proposition}
\begin{proof}
Let $(X_\epsilon,Y_\epsilon)$ be defined by the stochastic matrix
\begin{equation}
\P_{X,Y}=\begin{pmatrix}
\frac{1-\epsilon}{N} & \frac{\epsilon}{N} & 0  & \cdots & 0 \\
  0 & \frac{1-\epsilon}{N} & \frac{\epsilon}{N}  &  \cdots & 0\\
  \vdots & \vdots & \vdots & \ddots &  \ddots\\
  \frac{\epsilon}{N}  & 0 & \cdots & 0 & \frac{1-\epsilon}{N}  
 \end{pmatrix}. 
\end{equation}
Since there is a single class of communication, common information $X_\epsilon\land Y_\epsilon=0$ is zero for every $\epsilon >0$. By contrast, when $\epsilon=0$, $X_\epsilon\land Y_\epsilon$ is uniformly distributed among $N$ communication classes. Consequently, $D(X_\epsilon\land Y_\epsilon,0)=0$ for any $\epsilon>0$ whereas $D(X_0\land Y_0,0)=H(X_0\land Y_0)=\log N$ is arbitrarily large for $\epsilon=0$.
\end{proof}

However, it should be noted that the joint information $\vee$ is continuous with respect to Shannon's distance. In fact we have the following
\begin{Proposition} \label{continuity-total-info}
For any $X$, $X'$, $Y$, $Y'$,
\begin{equation}
D(X\vee Y, X'\vee Y') \leq D(X,X') + D(Y,Y'). 
\end{equation}
\end{Proposition}

\begin{proof}
One has
\begin{equation}
\begin{aligned}
  H(X\vee Y|X'\vee Y') &= H(X,Y|X',Y')\\
  &\overset{(a)}{=} H(X|X',Y') + H(Y|X',Y',X) \\
  &\overset{(b)}{\leq} H(X|X') + H(Y|Y').
\end{aligned} 
\end{equation}
where $(a)$ is the consequence of the chain rule and $(b)$ is due to the fact that conditioning reduces entropy.
Since $X$, $X'$ and $Y$, $Y'$ play a symmetrical role in $(b)$, we can permute the roles of $X$, $X'$ and $Y$, $Y'$, which gives $H(X'\vee Y'|X\vee Y) \leq H(X'|X) + H(Y'|Y)$. Summing both inequalities yields the result.
\end{proof}

\begin{Remark}
In particular for $X=X'$, for any $X,Y,Z$,
\begin{equation}
D(X\vee Y, X\vee Z) \leq D(Y,Z).
\end{equation}
In other words joining the same $X$ can only reduce the Shannon distance: In this respect, the joining operator $Y\mapsto X\vee Y$ is a contraction operator.
\end{Remark}

%

Furthermore, the entropy, the conditional entropy, and the mutual information are continuous with respect to the entropic distance of Shannon. Indeed, we have the following inequalities~(see Problem 3.5 in~\cite{CsiszarKorner11}):
\begin{Proposition}
For all $X$, $Y$, $X'$, $Y'$,
\begin{enumerate}
\item[\crm{1})] $|H(X) - H(Y)| \leq D(X,Y)$.
\item[\crm{2})] $|H(X,Y) - H(X',Y')| \leq D(X,X') + D(Y,Y')$.
\item[\crm{3})] $|H(X|Y) - H(X'|Y')| \leq D(X,X') + 2D(Y,Y')$. 
\item[\crm{4})] $|I(X;Y) - I(X';Y')| \leq 2(D(X,X')+D(Y,Y'))$.
\end{enumerate}
\end{Proposition}

\begin{proof}

\mbox{}

\begin{enumerate}
\item[\crm{1})] By the chain rule: $H(X) + H(Y|X) =H(X,Y)= H(Y) + H(X|Y)$, hence  $|H(X) - H(Y)| =|H(X|Y) - H(Y|X)| \leq H(X|Y) + H(Y|X) = D(X,Y)$.
\item[\crm{2})] 
Applying inequality \crm{1})~to the variables $(X,Y)$ and $(X',Y')$, we obtain $|H(X,Y) - H(X',Y')| \leq D((X,Y),(X',Y'))$.
From the continuity of joint information (Proposition~\ref{continuity-total-info}), one can further bound $D((X,Y),(X',Y')) \leq D(X,X') + D(Y,Y')$. 
\item[\crm{3})] 
By the chain rule, 
$|H(X|Y) - H(X'|Y')| = |H(X,Y) - H(Y) - (H(X',Y') - H(Y'))| \leq |H(X,Y) - H(X',Y')| + |H(Y') - H(Y)|$. 
The conclusion now follows from~\crm{1})~and~\crm{2}).
\item[\crm{4})] By the chain rule, 
$|I(X;Y) - I(X';Y')| = |H(X) - H(X') + H(Y) - H(Y') + H(X',Y') - H(X,Y)|
\leq |H(X) - H(X')| + |H(Y) - H(Y')| + |H(X',Y') - H(X,Y)|$. 
The conclusion follows from bounding
each of the three terms in the sum using \crm{1})~and~\crm{2}). \qedhere
\end{enumerate}
\end{proof}

In the remainder of this paper, we only consider quantities that are \emph{continuous} with respect to the entropic metrics (Shannon and Rajski distance). As a result, the discontinuity of the $\wedge$ operator will not hinder our derivations in the sequel.

\section{Geometric Properties of the Information Lattice} \label{geometry}


\subsection{Alignments of Random Variables} \label{alignment}

\begin{Definition}[Alignment]
Let $\delta$ be any distance on the information lattice. The random variables $X$, $Y$ and $Z$ are said to be \emph{aligned} with respect to $\delta$ if the triangular inequality is met with equality: 
\begin{equation}
\delta(X,Y) + \delta(Y,Z) = \delta(X,Z).
\end{equation}
\end{Definition}

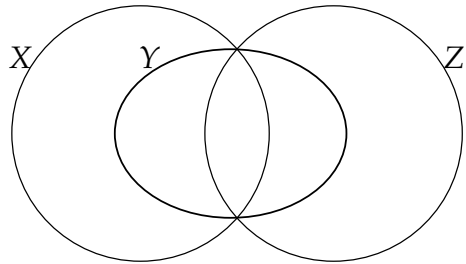
\begin{figure}[!ht]
\centering
\resizebox{.45\textwidth}{!}{%
\begin{circuitikz}
\tikzstyle{every node}=[font=\LARGE]
\draw[line width=.7pt ]  (-2.5,12.5) circle (2.5cm);
\draw[line width=.7pt ]  (1.25,12.5) circle (2.5cm);
\draw [line width=1pt ] (-0.75,12.5) ellipse (2.25cm and 1.65cm);
\node [font=\LARGE] at (-2.32,14) {$Y$};
\node [font=\LARGE] at (-4.84,14) {$X$};
\node [font=\LARGE] at (3.59,14) {$Z$};
\end{circuitikz}
}%
\caption{Venn diagram illustrating the alignment condition for the Shannon distance.}
\label{fig:alignD}
\end{figure}

\begin{Proposition}[Alignment w.r.t. the Shannon distance $D$]\label{alignment-Shannon}
The random variables $X$, $Y$ and $Z$ are aligned w.r.t. $D$ if and only if $X - Y - Z$ is a Markov chain and $Y \leq X\vee Z$. 
\end{Proposition}
This alignment condition is illustrated in Fig.~\ref{fig:alignD}.

\begin{proof}
From the proof of the triangular inequality for $D$ (Proposition~\ref{prop-D}),
equality holds iff equality holds in both inequalities $H(X|Z) \leq H(X,Y|Z) = H(X|Y,Z) + H(Y|Z) \leq H(X|Y) + H(Y|Z)$ and those inequalities obtained by permuting the roles of $X$ and $Z$.
Since $H(X,Y|Z) - H(X|Z) = H(Y|X,Z)$ and $H(X|Y) - H(X|Y,Z) = I(X;Z|Y)$, equality holds iff
$H(Y|X,Z) = 0$ and $I(X;Z|Y) = 0$, both conditions being symmetric in $(X,Z)$. 
Now $H(Y|X,Z) = 0$ means that $Y$ is a function of $(X,Z)$, i.e., $Y \leq X\vee Z$. Also $I(X;Z|Y) = 0$ means that $X$ and $Z$ are conditionally independent given $Y$, which characterizes the fact that $X - Y - Z$ forms a Markov chain.
\end{proof}

\begin{figure}[!ht]
\centering
    \resizebox{0.45\textwidth}{!}{%
    \begin{circuitikz}
    \tikzstyle{every node}=[font=\LARGE]
    \draw [line width=1pt ] (5,11.25) circle (3.75cm);
    \draw [line width=1pt ] (10,11.25) circle (3.75cm);
    \node [font=\LARGE] at (2,6.25) {$X$};
    \node [font=\LARGE] at (13.25,6.25) {$Z$};
    \node [font=\LARGE] at (7.5,16.5) {$Y$};
    \draw [ line width=0.8pt, -Stealth] (2.25,7) -- (3,8);
    \draw [ line width=0.8pt, -Stealth] (13,7) -- (12.25,8);
    \draw [ line width=0.8pt, -Stealth] (7.5,16.25) -- (8.75,15);
    \draw [ line width=0.8pt, -Stealth] (7.5,16.25) -- (6.25,15);
    \end{circuitikz}
    }%
\caption{Venn diagram illustrating the alignment condition for the Rajski distance.}
\label{fig:alignd}
\end{figure}
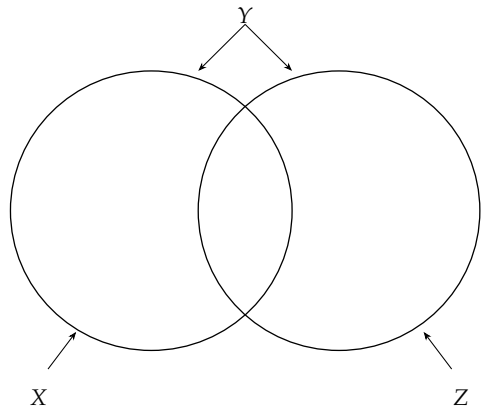

\begin{Proposition}[Alignment w.r.t. Rajski's distance $d$] \label{alignment_rajski}
The random variables $X$, $Y$ and $Z$ are aligned w.r.t. $d$ if and only if $Y = X\vee Z$. 
\end{Proposition}
This alignment condition is illustrated in Fig.~\ref{fig:alignd}.

\begin{proof}
From the proof of the triangular inequality for $d$ (Proposition~\ref{distance-Rajski}), equality holds iff equality holds in all inequalities \eqref{eq-rajski-denom-1}, \eqref{eq-rajski-denom-2}, and \eqref{demo-ineq-triangulaire-2}, and those inequalities obtained by permuting the roles of $X$ and $Z$.
Now a close inspection of~\eqref{eq-rajski-denom-1} shows it achieves equality iff $H(Z|Y) = 0$, that is, $Z \leq Y$.
Similarly, \eqref{eq-rajski-denom-2} achieves equality iff $H(X|Y) = 0$, that is $X \leq Y$.
Both conditions write $X\vee Z\leq Y$, which is symmetric in $(X,Z)$.
Finally, \eqref{demo-ineq-triangulaire-2} achieves equality iff $H(X|Y) + H(Y|Z) = H(X|Z)$ and the corresponding equality obtained by permuting the roles of $X$ and $Z$. This means that $X$, $Y$ and $Z$ are aligned w.r.t. $D$, that is, $X - Y - Z$ is a Markov chain and $Y \leq X\vee Z$. 
Overall $Y=X\vee Z$, which already implies that $X$ and $Z$ are conditionally independent given $Y=(X,Z)$, i.e., $X - Y - Z$ is a Markov chain.
\end{proof}

\begin{Remark}
Note that if $X$, $Y$ and $Z$ are aligned in the sense of Rajski's distance, then they are also aligned in the sense of Shannon's entropic distance since $Y = (X,Z)$ implies that $X - Y - Z$ is a Markov chain. 
Thus, the alignment condition is stronger in the case of the Rajski distance. 
\end{Remark}

\begin{Remark}
The alignment condition w.r.t. the Rajski distance is simpler and expressed  
by using only the operators of the information lattice, whereas that w.r.t. the Shannon distance requires the additional notion of Markov chain.
Therefore, in the sequel, we 
develop some geometrical aspects of the information lattice based essentially on the Rajski distance.
\end{Remark}


\subsection{Convex Sets of Random Variables in the Information Lattice}

\begin{Definition}[Convexity]
Given two random variables $X$ and $Y$, we define the \emph{segment} $[X,Y]$ of endpoints $X$ and $Y$ as the set of all random variables $Z$ such that $X$, $Z$ and $Y$ are aligned with respect to the Rajski distance, i.e., such that $d(X,Z) + d(Z,Y) = d(X,Y)$.

A set $\mathcal{C}$ of points (random variables) in the information lattice is \emph{convex} if for all points $X,Y \in \mathcal{C}$, the segment $[X,Y] \subseteq \mathcal{C}$. 
If $\mathcal{S}$ is any set of points of the information lattice, its \emph{convex envelope} is the smallest convex set containing $\mathcal{S}$.
\end{Definition}
By its very definition, the convex envelope of the two-element set $\{X,Y\}$ is the segment $[X,Y]$.
We have the following simple characterization.

\begin{Proposition}[Segment Characterization]
For any  two elements $X$, $Y$ of the information lattice, the segment $[X,Y]$ is the three-element set $[X,Y]= \bigl\{X,(X,Y),Y\bigr\}$, with respective distances to endpoints given by
$d\bigl(X,(X,Y)\bigr) = \frac{H(Y|X)}{H(X,Y)}$
and $d\bigl(Y,(X,Y)\bigr) = \frac{H(X|Y)}{H(X,Y)}$.
\end{Proposition}

\begin{proof}
$X$ and $Y$ do belong to the segment $[X,Y]$ since $d(X,X) + d(X,Y) = d(X,Y)$ and $d(X,Y) + d(Y,Y) = d(X,Y)$. Moreover, if $Z \in [X,Y]$, then $X$, $Z$ and $Y$ are aligned with respect to the Rajski distance so that necessarily $Z = (X,Y)$.
One calculates
$d\bigl(X,(X,Y)\bigr) = \frac{H(X|X,Y) + H(X,Y|X)}{H(X,Y)} = \frac{H(Y|X)}{H(X,Y)}$, and similarly for $d\bigl(Y,(X,Y)\bigr)$ by permuting the roles of $X$ and~$Y$.
\end{proof}

\begin{figure}[h!]
    \centering
    \begin{subfigure}{0.3\textwidth}
        \centering
        \begin{tikzpicture}
            \draw (1,1) node{$\bullet$};
            \draw (1,1) node[below left]{$X$};
            
            \draw (3.25,3.25) node{$\bullet$};
            \draw (3.25,3.25) node[above left]{$X \vee Y$};
            
            \draw (4,4) node{$\bullet$};
            \draw (4,4) node[above left]{$Y$};
            
            \draw[black] (1,1) [stealth-stealth] -- node[below right]{$\frac{H(Y|X)}{H(X,Y)}$} (3.25,3.25);
            \draw[black] (3.25,3.25) [stealth-stealth] -- node[below right]{$\frac{H(X|Y)}{H(X,Y)}$} (4,4);
        \end{tikzpicture}
        \caption{Arbitrary $X$ and $Y$}
    \end{subfigure}
    \begin{subfigure}{0.3\textwidth}
        \centering
            \begin{tikzpicture}
            \draw (1,1) node{$\bullet$};
            \draw (1,1) node[below left]{$X$};
            \draw (4,4) node{$\bullet$};
            \draw (4,4) node[above left]{$Y = X\vee Y$};
            \draw[black] (1,1) [stealth-stealth] -- node[below right]{$\frac{H(Y|X)}{H(Y)}$} (4,4) ;
        \end{tikzpicture}
        \caption{$X \leq Y$}
    \end{subfigure}
    \begin{subfigure}{0.3\textwidth}
        \centering
            \begin{tikzpicture}
                \draw (1,1) node{$\bullet$};
                \draw (1,1) node[above]{$X = Y = X \vee Y$};
            \end{tikzpicture}
        \caption{$X = Y$}
    \end{subfigure}
    \caption{Visualization of the segment $[X,Y]$ for three possible cases.}
    \label{fig:segment}
\end{figure}
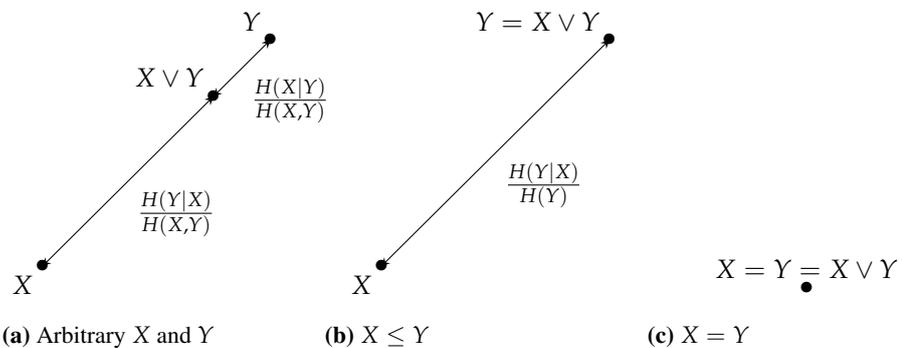

\begin{Remark}
By the above Proposition, segments in the information lattice are intrinsically \emph{discrete} objects. 
In the case where $X \leq Y$ or $Y \leq X$, then the segment $[X,Y]$ contains only two distinct points, $X$ and $Y$.
Obviously, if $X = Y$, then $[X,Y]$ is a singleton.
This gives three possible cases as illustrated in Fig.~\ref{fig:segment}.
\end{Remark}

\begin{Remark}
As a result of this characterization, four or more distinct points cannot be aligned w.r.t. the Rajski distance, because a segment cannot contain more than three distinct points.  
\end{Remark}

\begin{Proposition}\label{prop-characconvex}
$\mathcal{C}$ is convex iff it is closed under the $\vee$ operator.
\end{Proposition}

\begin{proof}
$\mathcal{C}$ is convex iff for all $X,Y \in \mathcal{C}$, $[X,Y] \subseteq \mathcal{C}$, that is, $X$, $Y$ and $(X,Y) = X \vee Y \in  \mathcal{C}$.
\end{proof}

Beyond the case of a two-element set, we now characterize the convex envelope of any $n$-element set in the information lattice, that is, the convex envelope of $n$ random variables $X_{1}$, $X_{2}$,..., $X_{n}$.
We adopt the following usual convention.
For any $n$-tuple of indices $I = (i_{1},i_{2},...,i_{n})$, the random vector $(X_{i_{1}},X_{i_{2}},...,X_{i_{n}})=X_{i_{1}}\vee X_{i_{2}}\vee \cdots \vee X_{i_{n}}$ is denoted by 
$X_{I}$. Again by convention, for the empty set, $X_{\emptyset} = 0$, so that one always have $X_{I\cup J}=X_I\vee X_J$ for any two finite sets of indices $I$ and $J$.

\begin{Proposition}\label{prop-convexenvelope}
Let $I$ be a finite index set and $(X_{i})_{i \in I}$ be random variables. The convex envelope of $(X_{i})_{i \in I}$ is $ \{ \vee_{j \in J}X_{j} \mid \emptyset\not=J \subseteq I \}=
\{ X_{J} \mid \emptyset\not=J \subseteq I  \}$, that is, the set of all sub-tuples of the $X_i$.
\end{Proposition}

\begin{proof}
With every $X_i$ ($i \in I$), the convex envelope in question should be closed by the $\vee$ operator, hence contain any tuple $\vee_{j \in J}X_{j}$ for any nonempty $J\subseteq I$. Now
$\mathcal{C} = \{  \vee_{j \in J}X_{j}=X_{J} \mid \emptyset\not=J \subseteq I  \}$ contains all $X_{i}$ for $i \in I$ and is already convex. Indeed, for all $X_{J} \in \mathcal{C}$ and $X_{K} \in \mathcal{C}$, $X_{J} \vee X_{K} = X_{J \cup K} \in \mathcal{C}$.
%
\end{proof}


\begin{Remark}
Given a finite set $I$ of index of cardinality $|I| =n$, the convex envelope of $(X_{i})_{i \in I}$ contains at most $2^{n} - 1$ distinct elements, since there are $2^{n}-1$ nonempty subsets of $I$. The number $2^n-1$ is only an upper bound since it might happen that two different subsets $J$ and $K$ of $I$ are such that $X_{J} = X_{K}$. 
\end{Remark}

\begin{figure}[!ht]
    \begin{center}
        \resizebox{0.4\textwidth}{!}{%
        \begin{tikzpicture}
            \tikzstyle{every node}=[font=\footnotesize]
        
        \draw (0,0) node{$\bullet$};
        \draw (0,0) node[left]{$X_{1}$};
        
        \draw (1,1.732) node{$\bullet$};
        \draw (1,1.732) node[above]{$X_{0}$};
        
        \draw (2,0) node{$\bullet$};
        \draw (2,0) node[right]{$X_{2}$};
        
        \draw (1,0) node{$\bullet$};
        \draw (1,0) node[below]{$X_{1} \vee X_{2}$};
        
        \draw (0.5,0.866) node{$\bullet$};
        \draw (0.5,0.866) node[above left]{$X_{0} \vee X_{1}$};
        
        \draw (1.5,0.866) node{$\bullet$};
        \draw (1.5,0.866) node[above right]{$X_{0} \vee X_{2}$};
        
        \draw (1,0.6666) node{$\bullet$};
        \draw (1,0.6666) node[below]{$X_{0} \vee X_{1} \vee X_{2}$};
            
        \end{tikzpicture}
        }
    \end{center}
    \caption{$7$-element convex envelope of three random variables $X_{0}$, $X_{1}$ and $X_{2}$. These three random variables are represented as vertices of an (equilateral) triangle. The other points in the convex envelope are obtained as intersections of medians and edges, and the common information $X_{0}\vee  X_{1}\vee X_{2}$ is the center of gravity (intersection of the three medians).
    Similarly, the 15-element convex envelope of four distinct points can be visualized in a tetrahedron, etc.}
    \label{fig:conv-hull}
\end{figure}
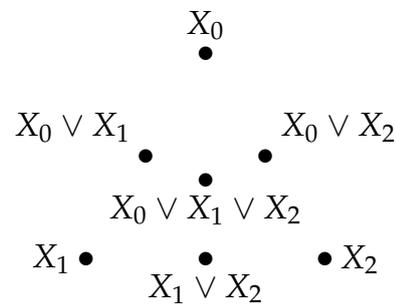

An example of the convex envelope of a family of three random variables is shwon in Fig.~\ref{fig:conv-hull}.

It is also interesting to note that any sublattice of the information lattice does have some convexity properties:
\begin{Proposition}\label{prop-latticeconvex}
    Any sublattice of the information lattice (including the information lattice itself) is \emph{convex}.
    With every $X_i$ ($i \in I$), the sublattice also contains the convex envelope of $(X_{i})_{i \in I}$.
    \label{prop:sub-lat}
\end{Proposition}

%
%
%

\begin{proof}
With every two points $X,Y$, the sublattice should contain their maximum $X\vee Y$, hence the whole segment $[X,Y]$. It is, therefore, convex. Now every convex set contains the convex enveloppe of any of its subsets.
\end{proof}

\subsection{The Lattice Generated by a Random Variable}

In the sequel, we are interested in all possible deterministic functions of a given random variable~$X$. In fact, their set constitutes a sublattice of the information lattice:

\begin{Proposition}[Sublattice generated by a random variable]
Let $X$ be any random variable in the information lattice. The set of all random variables $\leq X$ is a sublattice, which we call \emph{lattice generated by $X$}, denoted $\langle X \rangle$. It is a bounded lattice with maximum (total information) $X$ and minimum $0$.
\end{Proposition}

\begin{proof}
Let $Y \leq X$ and $Z \leq X$. There exists deterministic functions $f$ and $g$ such that $Y = f(X)$ a.s. and $Z = g(X)$ a.s.  Clearly 
$Y \wedge Z \leq Y \leq X$ and $Y \vee Z = (Y,Z) = (f(X),g(X)) \leq X$.
Therefore, the set of random variables $\leq X$ forms a sublattice. Clearly $X$ is maximum and $0$ (deterministic random variable seen as a constant function of $X$) is minimum.
\end{proof}

\begin{Remark}
One may also define the sublattice $\langle X_1,X_2,\ldots,X_n\rangle$ generated by several random variables $X_1,X_2,\ldots,X_n$ simply as the sublattice generated by the variable $X_1\vee X_2 \vee \cdots\vee X_n$. Therefore, it is enough to restrict ourselves to one random variable $X$ as the lattice generator.
\end{Remark}

\begin{Proposition}
$\langle X \rangle$ 
 is a complemented lattice.
\end{Proposition}

\begin{proof}
Let $Y\leq Z\leq X$, so that both $Y,Z \in \langle X \rangle$. By Proposition~\ref{prop-complemented}, $Y$ admits at least one complement information $\overline{Y}$ w.r.t. $Z$ in the information lattice, such that 
$Y \wedge \overline{Y} = 0$ and $Y \vee \overline{Y} = Z$. 
Now $\overline{Y} \leq Z \leq X$, hence the complement $\overline{Y} \in \langle X \rangle$ belongs to the sublattice generated by $X$. 
\end{proof}

\subsection{Properties of Rajski and Shannon Distances in the Lattice Generated by a Random Variable}

We now investigate the metric properties of the sublattice $\langle X\rangle$ generated by a random variable~$X$. 
To avoid the trivial case $\langle 0 \rangle = \{0\}$ we assume that $X$ is \emph{nondeterministic}.
First of all, we observe that the entropy of a element of the sublattice increases as it is closer to $X$ (in terms of either Shannon's or Rajski's distance):

%
%
%
%


\begin{Proposition}\label{prop-dist<X>}
For any $Y \in \langle X \rangle$, one has 
$D(X,Y)=H(X|Y)=H(X)-H(Y)$ and
$d(X,Y) 
= \dfrac{H(X|Y)}{H(X)} = 1 - \dfrac{H(Y)}{H(X)}$.
In particular maximum distance $d(X,Y) =1$ is achieved iff $Y=0$.
\end{Proposition}

\begin{proof} One has
\begin{equation}
\begin{aligned}
d(X,Y) &= \frac{D(X,Y)}{H(X,Y)}= \frac{H(X|Y) + H(Y|X)}{H(X)} \\
&\overset{(a)}{=} \frac{H(X|Y)}{H(X)} 
\overset{(b)}{=} \frac{H(X) - H(Y)}{H(X)} 
= 1 - \frac{H(Y)}{H(X)}.
\end{aligned} 
\end{equation}
where $(a)$ is because $Y \leq X$, and $(b)$ is a consequence of the chain rule: $H(X) = H(X,Y) = H(Y) + H(X|Y)$. 
\end{proof}


\begin{Remark}\label{rmk-redundancy}
In the language of data compression, $d(X,Y) 
= \dfrac{H(X)-H(Y)}{H(X)} = 1 - \dfrac{H(Y)}{H(X)}$ can be seen as the relative entropic \emph{redundancy} of $X$ when it is represented (``encoded'') by $Y$.
\end{Remark}

\begin{Remark}
The maximum distance case in the Proposition can be stated as follows: The only random variables $Y$ that can be obtained as functions of $X$ ($Y\in\langle X \rangle$) while being also independent of $X$ ($d(X,Y)=1$) are the constant (deterministic) random variables.
\end{Remark}

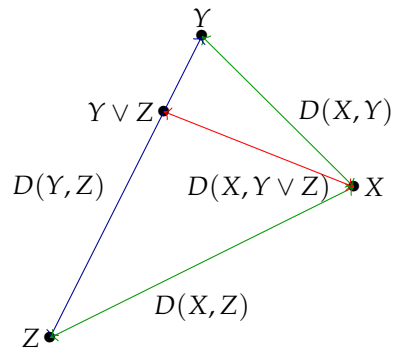
\begin{figure}[h!]
    \begin{center}
        \begin{tikzpicture}
            
        \coordinate[label=right:$X$] (X) at (4,2);
        \coordinate[label=above:$Y$] (Y) at (2,4);
        \coordinate[label=left:$Z$] (Z) at (0,0);
        \coordinate[label=left:$Y\vee Z$] (D) at (1.5,3);
        
        \draw (4,2) node{$\bullet$};
        \draw (2,4) node{$\bullet$};
        \draw (0,0) node{$\bullet$};
        \draw (1.5,3) node{$\bullet$};
                        
        \draw[green] (X) [<->] -- (Y);
        \draw[blue] (Y) [<->] -- (Z);
        \draw[green] (Z) [<->] -- (X);
        \draw[red] (X) [<->] -- (D);
        
        \tkzLabelSegment[below=8pt](X,Z){$D(X,Z)$}
        \tkzLabelSegment[left=4pt](Y,Z){$D(Y,Z)$}
        \tkzLabelSegment[right=4pt](Y,X){$D(X,Y)$}
        \tkzLabelSegment[below=6pt](D,X){$D(X,Y\vee Z)$}
        
        \end{tikzpicture}
    \end{center}
    \caption{Graphical representation of Apollonius's theorem (Lemma \ref{lemme2}).}
    \label{fig:lemma-2}
\end{figure}

In Euclidean geometry, \emph{Apollonius's theorem} allows one to calculate the length of the median of a triangle $XYZ$ given the length of its other three sides. In the information lattice context, $Y\vee Z$ denotes the median of the segment $[Y,Z]$ (the only possible point in the segment that is not an endpoint). Thus, Apollonius's theorem gives a formula for distance $D(X,Y\vee Z)$ in terms of $D(X,Y)$, $D(X,Z)$, and $D(Y,Z)$.
The following Proposition is the analogue of {Apollonius's theorem} for the Shannon distance in the information lattice generated by $X$:

\begin{Lemma}[Apollonius's theorem in $\langle X\rangle$] \label{lemme2}
For any $Y,Z\in\langle X \rangle$,
\begin{equation}
 D(X,Y\vee Z) = \frac{D(X,Y) + D(X,Z) - D(Y,Z)}{2}.
\end{equation}
This can also be written as
\begin{equation}
D(X,Y)+D(X,Z)=D(Y,Z) + 2 D(X,Y\vee Z) 
\end{equation}
\end{Lemma}
This is illustrated in Fig.~\ref{fig:lemma-2}.
Note that when $X=Y\vee Z$ one recovers that $Y,X,Z$ (in this order) are aligned.

\begin{proof}
From Proposition~\ref{prop-dist<X>}, $D(X,Y)=H(X)-H(Y)$ for any $Y\in\langle X \rangle$, in particular 
$D(X,Z)=H(X)-H(Z)$ and $D(X,Y\vee Z)=H(X)-H(Y,Z)$ also.
%
%
%
Therefore, $D(X,Y)+D(X,Z)- 2 D(X,Y\vee Z) = 2H(Y,Z)-H(Y)-H(Z)=H(Z|Y)+H(Y|Z)=D(Y,Z)$.
%
%
%
%
%
%
%
%
\end{proof}


From Lemma \ref{lemme2} we derive the following 
\begin{Lemma}\label{lemme4}
For any $Y,Z\in\langle X \rangle$,
\begin{equation}
d(X,Y)+d(X,Z) \leq d(X,Y\vee Z) +1
\end{equation}
with equality if and only if $Y$ and $Z$ are independent.
\end{Lemma}

\begin{proof}
Observe that $D(Y,Z) + D(X,Y\vee Z)=H(Y|Z)+H(Z|Y) + H(X|Y\vee Z)\leq H(Y)+H(Z|Y) + H(X|Y,Z)=H(Y,Z) + H(X|Y,Z)=H(X,Y,Z)=H(X)$ since $Y,Z\in\langle X \rangle$, with equality iff $Y$ and $Z$ are independent.
Now by Lemma~\ref{lemme2}, $D(X,Y)+D(X,Z)=D(Y,Z)+2$\hbox{$D(X,Y\vee Z)\leq D(X,Y\vee Z)$}$+H(X)$.
Dividing by $H(X)=H(X,Y)=H(X,Z)=H(X,Y,Z)$ yields the announced inequality.
%
%
%
\end{proof}

In the other direction we have the following
\begin{Lemma}\label{lemme5}
For any $Y,Z\in\langle X \rangle$,
\begin{equation}
 d(X,Y\vee Z) \leq d(X,Y) + d(X,Z)
\end{equation}
with equality if and only if $X=Y=Z$. 
\end{Lemma}

\begin{proof}
By the triangular inequality, $d(X,Y\vee Z)\leq d(X,Y)+d(Y,Y\vee Z)$ with equality iff $Y=X\vee Y\vee Z=X$ by the alignment condition. Similarly, $d(X,Y\vee Z)\leq d(X,Z)+d(Z,Y\vee Z)$ with equality iff $Z=X\vee Y\vee Z=X$. Summing the two inequalities, $2d(X,Y\vee Z)\leq d(X,Y)+d(X,Z) + d(Y,Y\vee Z)+d(Z,Y\vee Z)$ where 
$d(Y,Y\vee Z)+d(Z,Y\vee Z)=d(Y,Z)\leq d(X,Y)+d(X,Z)$ with equality iff $X=Y\vee Z$. Combining yields the announced inequality.
%
%
%
%
%
%
%
\end{proof}

\begin{Remark}
In the course of the proof, we have proved the following stronger inequality: for any $Y,Z\in\langle X \rangle$,
\begin{equation}
d(X,Y\vee Z) \leq \frac{d(X,Y)+d(Y,Z) +d(Z,X)}{2}
\end{equation}
with the same equality condition $X=Y=Z$.
\end{Remark}

\begin{Remark}
By Lemmas~\ref{lemme4} and~\ref{lemme5}, we see that in terms of Rajski distances to the generator $X$, $d(X,Y\vee Z)$ lies between $d(X,Y)+d(X,Z)-1$ and $d(X,Y)+d(X,Z)$, where the lower and upper bounds differ by $1$ and the minimum value is achieved in the case of independence. These two Lemmas are instrumental in the derivations of the next section.
\end{Remark}

\section{The Perfect Reconstruction Problem} \label{reconstruction}

\subsection{Problem Statement}



Suppose one is faced with the following reconstruction problem. We are given a (discrete) source of information $X$ (e.g., a digital signal, some text document, or any type of data), which is processed using deterministic functions into several ``components'':
\begin{equation}
X_1=f_1(X), \quad X_2=f_2(X),  \quad\ldots, \quad X_n=f_n(X)
\end{equation}
(e.g., different filtered versions of the signal at various frequencies, translated parts of the document, or some nonlinear transformations of the data). The natural question is: Did one \emph{loose information} when processing $X$ into its $n$ components $X_1,X_2,\ldots,X_n$? Or else, can we \emph{perfectly reconstruct} the original $X$ from its $n$ components using some (unknown) deterministic function $X=f(X_1,\ldots,X_n)$? 

We emphasize that all involved functions must be \emph{deterministic} (no noise is involved), otherwise \emph{perfect} reconstruction (without error) would not be possible. Yet we do not require any precise form for the reconstruction function $f$, only that such a reconstruction exists.
To our knowledge, the first occurence of such a problem (for $n=2$) is an Exercise~6 of the textbook~\cite{Rioul07}.

Stated in the information lattice language, the perfect reconstruction problem is as follows. Suppose we are given $X_1,X_2,\ldots,X_n$ in $\langle X\rangle$, the sublattice generated by $X$. Is it true that $X\leq X_1\vee X_2\vee\cdots\vee X_n$? Since the sublattice is convex (Proposition~\ref{prop-latticeconvex}), i.e., stable by the $\vee$ operator (Proposition~\ref{prop-characconvex}), one always has, by assumption, that $X_1\vee X_2\vee\cdots\vee X_n\in \langle X\rangle$, i.e., $X_1\vee X_2\vee\cdots\vee X_n\leq X$. Therefore, in the reconstruction problem, it is equivalent to determine whether $X= X_1\vee X_2\vee\cdots\vee X_n$ or $X\not= X_1\vee X_2\vee\cdots\vee X_n$.

\begin{Remark}\label{rmk-triangle}
Geometrically, by Proposition~\ref{prop-convexenvelope}, determining whether $X\leq X_1\vee X_2\vee\cdots\vee X_n$ or not is equivalent to determining whether $X$ is in the \emph{convex envelope} of the $(X_{i})_{i =1,\ldots,n}$. 

Thus, when $n=2$, perfect reconstruction is possible iff $X$ lies in the segment $[X_1,X_2]$.
When $n=3$, perfect reconstruction is possible iff for every distinct indices 
$i,j,k\in\{1,2,3\}$, $X_{i}$, $X$ and $X_{j,k}$ are aligned w.r.t. the Rajski distance as illustrated in Fig.~\ref{fig:three-recon}.

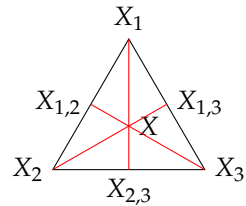
\begin{figure}[!ht]
    \begin{center}
        \begin{tikzpicture}
        
        \coordinate[label=left:$X_{2}$] (A) at (0,0);
        \coordinate[label=above:$X_{1}$] (B) at (1,1.732051);
        \coordinate[label=right:$X_{3}$] (C) at (2,0);
        
        \coordinate[label=below:$X_{2,3}$] (D) at (1,0);
        \coordinate[label=left:$X_{1,2}$] (E) at (0.5,0.86603);
        \coordinate[label=right:$X_{1,3}$] (F) at (1.5,0.86603);
        
        \coordinate[label=right:$X$] (G) at (1,0.57735);
        
        \coordinate (H) at (1.01,0.57);
        \coordinate (I) at (0.99,0.57);
        \coordinate (J) at (1,0.58);
        
        \draw (A) -- (B) -- (C) -- cycle;
        \draw[red] (C) -- (E);
        \draw[red] (A) -- (F);
        \draw[red] (B) -- (D);
        
        \end{tikzpicture}
    \end{center}
    \caption{Geometric Illustration of the Three-Component Reconstruction Problem.}
    \label{fig:three-recon}
\end{figure}
 
\end{Remark}

Intuitively, the processed components $X_i$ should not (on the whole) be too ``far away'' from the original source $X$ in order that perfect reconstruction be possible. In other words, at least some of the distances $d(X,X_i)$ should not be too high. Such distances can be, in principle, evaluated when processing the source $X$ into each of its components. In the following Subsection, we give a simple necessary condition on the sum $d(X,X_1)+d(X,X_2)+\cdots+d(X,X_n)$ to allow for perfect reconstruction.

\subsection{A Necessary Condition for Perfect Reconstruction}

The main result of this paper is the following
\begin{Theorem}[Necessary Condition for Perfect Reconstruction] 
\label{thm:necessary-cond}
Let $X$ be a random variable and let $X_1,X_2,\ldots,X_n\in\langle X\rangle$. If perfect reconstruction is possible: $X= X_1\vee X_2\vee\cdots\vee X_n$, then
\begin{equation}\label{eq-possible?}
\sum_{i=1}^{n}d(X,X_{i}) \leq n-1
\end{equation}
with equality iff  $X_1,X_2,\ldots,X_n$ are independent.
\end{Theorem}

\begin{proof}
By repeated use of Lemma~\ref{lemme4}, each joining operation of two components in the sum---e.g., passing from $d(X,X_i)+d(X,X_j)$ to $d(X,X_i\vee X_j)$---decreases this sum by at most $1$. Thus,
\begin{equation}
\begin{aligned}
 \sum_{i=1}^{n}d(X,X_{i}) &\leq  \sum_{i=1}^{n-2}d(X,X_{i})+ d(X,X_{n-1}\vee X_n) + 1 \\
 &\leq \sum_{i=1}^{n-3}d(X,X_{i})+ d(X,X_{n-2}\vee X_{n-1}\vee X_n) + 2\\
 &\;\;\vdots\\
 &\leq d(X,X_1\vee X_2\vee\cdots\vee X_n) + n-1 = n-1.
\end{aligned}
\end{equation}
Equality holds iff all the above $n-1$ inequalities are equalities. By the equality condition of Lemma~\ref{lemme4}, this means by induction that $X_1$ is independent from $X_2\vee \cdots \vee X_{n}$, where $X_2$ is independent from 
$X_3\vee \cdots \vee X_{n}$, and so on until $X_{n-1}$ is independent from $X_n$. Overall this is equivalent to saying that all components $X_1,X_2,\ldots,X_n$ are mutually independent.
\end{proof}

\begin{Remark}\label{rmk:pairwise}
To illustrate Theorem~\ref{thm:necessary-cond}, consider a uniformly distributed two-bit random variable $X$ (i.e., the result of two independent coin flips) and let $X_1$ be the result of the first coin toss and $X_2$ be that of the second coin toss. Clearly, reconstruction is possible since $X=(X_1,X_2)$. Now a simple calculation gives $d(X,X_1)=\frac{H(X|X_1)}{H(X)} = \frac{\log 2 }{ \log 4} =\frac{1}{2}$, and similarly $d(X,X_2)=\frac{1}{2}$, which shows that~\eqref{eq-possible?} is achieved with equality: $d(X,X_1)+d(X,X_2)=2-1=1$. This is not surprising since $X_1$ and $X_2$ are independent, as can be checked directly.

Now consider $X_3=0$ or $1$ depending on whether $X_1=X_2$ or not. Clearly, $X$ can be also reconstructed from $X_1,X_2,X_3$ since it can already be reconstructed from $X_1,X_2$. Again one computes $d(X,X_3)=\frac{\log 2 }{ \log 4} =\frac{1}{2}$ so in this case, the sum of distances to $X$ is now $d(X,X_1)+d(X,X_2)+d(X,X_3)=\frac{3}{2}<3-1=2$. This shows that~\eqref{eq-possible?} is still satisfied, but not with equality. In fact, it can easily be proved that even though $X_1,X_2,X_3$ are \emph{pairwise} independent, they are \emph{not mutually} independent. 
\end{Remark}

\begin{Remark}
In practice, Theorem~\ref{thm:necessary-cond} gives an \emph{impossibility} condition for perfect reconstruction of the random variable $X$ from components $X_1,X_2,\ldots,X_n$. Indeed, if the latter are such that 
\begin{equation}\label{eq-impossible}
\sum_{i=1}^{n}d(X,X_{i}) > n-1
\end{equation}
then perfect reconstruction is \emph{impossible}, however complex the reconstruction function $f$ could have been. In other words, $X< X_1\vee X_2\vee\cdots\vee X_n$, information was lost by processing.

That perfect reconstruction is impossible does not mean that it would never be possible to deduce one particular value of $X$ from some particular values of $X_1,X_2,\ldots,X_n$. It means that such a deduction is not possible in general, for every possible values taken by $X_1,X_2,\ldots,X_n$. In other words, there is at least one set of values $X_1=x_1$, $X_2=x_2$, \ldots, $X_n=x_n$ for which $X$ cannot be reconstructed unambiguously.
%
%
\end{Remark}

\begin{Remark}\label{rmk-rho}
Another look at Theorem~\ref{thm:necessary-cond}  can be made using the dependency coefficient $\rho=1-d$ in place of the Rajski distance.  Then the impossibility condition~\eqref{eq-impossible} simply writes
\begin{equation}\label{eq-imprho}
 \sum_{i=1}^{n}\rho(X,X_{i})<1.
\end{equation}
In other words, perfect reconstruction can only occur if the components are (as a whole) sufficiently dependent on the original $X$. Otherwise~\eqref{eq-imprho} precludes perfect reconstruction.
\end{Remark}

\begin{Remark}\label{rmk-sumvalue}
Since the Rajski distance is always upper bounded by $1$, if the impossibility condition~\eqref{eq-impossible} is met, then the actual value of the sum $\sum_{i=1}^{n}d(X,X_{i})$ necessarily lies in the interval $(n-1,n]$. 

In the worse situation $\sum_{i=1}^{n}d(X,X_{i})=n$,  all terms should equal one: $d(X,X_{i})=1$. This means that all components are independent from $X$. By Proposition~\ref{prop-dist<X>}, the components $X_i=0$ are all constants: In this case, all information is lost.
\end{Remark}

\begin{Remark}
By Theorem~\ref{thm:necessary-cond}, for perfect reconstruction to be possible, the components $X_i$ should be (at least slightly) tightened around $X$ in the sense that~\eqref{eq-possible?} is satisfied.
The example of Remark~\ref{rmk:pairwise} shows that under this condition (even when the inequality is strict), it may be actually possible to reconstruct $X$. 
However, proximity may not be enough: The necessary condition of Theorem~\ref{thm:necessary-cond} is \emph{not} sufficient in general.
 
To see this, consider $X$ uniformly distributed in the integer interval $\{0,1,\ldots,11\}$ and define $X_1=k$ if $X=2k$ or $2k+1$ and $X_2=\ell$ if $X=3\ell$, $3\ell+1$ or $3\ell+2$. In other words $X_1$ is the integer division of $X$ by $2$ and $X_2$ is the integer division of $X$ by $3$. One easily computes
\begin{equation}
d(X,X_1)+d(X,X_2)=\frac{H(X|X_1)+H(X|X_2)}{H(X) }= \frac{\log 2+\log 3}{\log 12}= \frac{\log 6}{\log 12}<1. 
\end{equation}
While the necessary condition~\eqref{eq-possible?} of Theorem~\ref{thm:necessary-cond}  is met, the value of $X$ cannot be unambiguously determined from those of $X_1$ and $X_2$. For example, $X_1=X_2=0$ leaves two possibilities $X=0$ or~$1$. Therefore, perfect reconstruction is not possible.

Another way to see this is to observe that perfect reconstruction is equivalent to saying that $X_1,X,X_2$ are aligned, which in terms of the Shannon distance would write $D(X_{1},X_{2}) = D(X,X_{1}) + D(X,X_{2})$.
But while $D(X,X_{1}) + D(X,X_{2})=\log 6$, one has 
\begin{equation}
D(X_1,X_2)= H(X_1|X_2)+H(X_2|X_1)= \bigl( \frac{1}{3}\log 3 + \frac{2}{3}\log \frac{3}{2}\bigr)+ \frac{2}{6}\log 2=\log 3 - \frac{\log 2}{3}
\end{equation}
which is clearly less than $\log 6$.
Therefore, perfect reconstruction is impossible in our example, because $X_{1}$ and $X_{2}$ are too close together, i.e., there is too much redundant information between them.

A slight modification of the above example where $X$ takes values in $\{0, 1, \ldots, 12m-1\}$ for abitraily large $m$ shows that the sum $d(X,X_1)+d(X,X_2)=\frac{\log 6}{\log(12m)}$ can actually be as small as desired, while perfect reconstruction is still impossible.  Therefore, there can be no condition of the form 
$\sum_{i=1}^{n}d(X,X_{i})<c$ (or any condition based only on the value of this sum) to ensure perfect reconstruction. Such a sufficient condition cannot be established without assuming some other property of the components $X_i$, as seen in the next Subsection.
\end{Remark}

\subsection{A Sufficient Condition for Perfect Reconstruction}

For independent components $X_1,X_2,\ldots,X_n$ (with no redundant information between them), the necessary condition of Theorem~\ref{thm:necessary-cond} becomes also a sufficient condition:


\begin{Theorem} [Sufficient Condition for Perfect Reconstruction] 
\label{thm:sufficient}
Let $X$ be a random variable and let $X_1,X_2,\ldots,X_n\in\langle X\rangle$ be \emph{independent}.
If inequality~\eqref{eq-possible?} holds, then it necessarily holds with equality:
\begin{equation}
\sum_{i=1}^{n}d(X,X_{i}) = n-1
\end{equation}
and perfect reconstruction is possible:  $X= X_1\vee X_2\vee\cdots\vee X_n$.
\end{Theorem}

\begin{proof}
A closer look at the proof of Theorem~\ref{thm:necessary-cond} shows that we have established (without the perfect reconstruction assumption) the general inequality
\begin{equation}\label{eq-general-cond}
 \sum_{i=1}^{n}d(X,X_{i}) \leq  d(X,X_1\vee X_2\vee\cdots\vee X_n) + n-1
\end{equation}
which holds with equality iff $X_1,X_2,\ldots,X_n$ are independent. Therefore, by the independence assumption,~\eqref{eq-possible?} writes $ \sum_{i=1}^{n}d(X,X_{i}) =  d(X,X_1\vee X_2\vee\cdots\vee X_n) + n-1\leq n-1$. Since the distance is nonnegative, this necessarily implies that the inequality holds with equality and that $d(X,X_1\vee X_2\vee\cdots\vee X_n)=0$, that is, $X= X_1\vee X_2\vee\cdots\vee X_n$.
\end{proof}

\begin{Remark}
Following Remark~\ref{rmk-sumvalue}, we see that for independent $X_1,X_2,\ldots,X_n$, the sum of distances to $X$: $\sum_{i=1}^{n}d(X,X_{i})$  can only take values in the interval $[n-1,n]$, with two possibilities:
\begin{itemize}
\item Either $\sum_{i=1}^{n}d(X,X_{i}) =n-1$, perfect reconstruction is possible;
\item or $\sum_{i=1}^{n}d(X,X_{i}) >n-1$, perfect reconstruction is impossible.
\end{itemize}
 In other words, independent components cannot be arbitrarily tightly packed around $X$.
 
 Following Remark~\ref{rmk-rho}, in terms of dependency coefficients, for independent $X_1,X_2,\ldots,X_n$,
\begin{itemize}
 \item Either $\sum_{i=1}^{n}\rho(X,X_{i}) =1$, perfect reconstruction is possible;
\item or $\sum_{i=1}^{n}\rho(X,X_{i}) <1$, perfect reconstruction is impossible.
\end{itemize}
\end{Remark}

%
%
%
%

\begin{Remark}
Following Remark~\ref{rmk-triangle} and Fig.~\ref{fig:three-recon}  in the case of three \emph{independent} components $X_{1},X_{2},X_{3}$, one should have $d(X,X_{1}) + d(X,X_{2}) + d(X,X_{3}) = 2$ for perfect reconstruction to hold. Incidentally, the graphical Euclidean illustration of Fig.~\ref{fig:three-recon} is faithful in this case, since for an equilateral triangle $X_{1}X_{2}X_{3}$ with sides of length 1, the sum of Euclidean distances equals $d(X,X_{1}) + d(X,X_{2}) + d(X,X_{3}) = \frac{2}{3} + \frac{2}{3} + \frac{2}{3}= 2$.
\end{Remark}

\subsection{Approximate Reconstruction}

Suppose we encode the information source $X$ by $n$ components $X_1,X_2,\ldots,X_n$ but do not particularly insist that \emph{perfect} reconstruction is possible. Rather, we assume that the encoding removes a \emph{fraction of redundancy} in $X$ equal to
\begin{equation}\label{eq-redundancy}
d(X,X_1\vee X_2\vee\cdots\vee X_n) = \delta
\end{equation}
(see Remark~\ref{rmk-redundancy}). Since the case $\delta=0$ corresponds to the previous case of perfect reconstruction ($X= X_1\vee X_2\vee\cdots\vee X_n$), we assume that $\delta>0$ in the sequel. Thus, in what follows, reconstruction of $X$ can only be approximate (up to a certain distance tolerance $\delta$). We then have the following

\begin{Theorem}[Approximate Reconstruction]\label{thm:approx}
Let $X$ be a random variable and let $X_1,X_2,\ldots,X_n\in\langle X\rangle$ such that~\eqref{eq-redundancy} holds with redundancy $=\delta>0$. Then
\begin{equation}\label{eq-delta}
\delta< \sum_{i=1}^{n}d(X,X_{i}) \leq n-1 +\delta. 
\end{equation}
with equality in the second inequality iff the components $X_1,X_2,\ldots,X_n$ are independent.
\end{Theorem}

\begin{proof}
The rightmost inequality in~\eqref{eq-delta} is just~\eqref{eq-general-cond} (with the announced case of equality), which was established by repeated application of Lemma~\ref{lemme4}. 
Similarly, repeated application of Lemma~\ref{lemme5} gives 
\begin{equation}
d(X,X_1\vee X_2\vee\cdots\vee X_n)\leq   \sum_{i=1}^{n}d(X,X_{i})
\end{equation}
with equality iff all $X_i=X$ ($i=1,\ldots,n$). But such an equality condition would yield $\delta=d(X,X)=0$, contrary to the assumption $\delta>0$. This shows the leftmost inequality in~\eqref{eq-delta} is strict.
\end{proof}

\begin{Remark}
Similarly as in the above two Subsections,  one can deduce from Theorem~\ref{thm:approx} that for independent components $X_1,X_2,\ldots,X_n$, one necessarily has $\sum_{i=1}^{n}d(X,X_{i}) = n-1 +\delta$, and that in general, approximate reconstruction within distance tolerance $\leq\delta$ will be impossible if $\sum_{i=1}^{n}d(X,X_{i}) > n-1 +\delta$.
\end{Remark}

\section{Examples and Applications} \label{application}

In this Section, we develop five examples of applications of the theorems of Section~\ref{reconstruction}. 

\subsection{Reconstruction from Sign and Absolute Value}\label{S-abssign}

Consider a real-valued random variable $X$ and assume that it is symmetric ($X$ is identically distributed as $-X$) and that $\P(X=0)=0$.  
 Now define $X_{1} = |X|$ (absolute value) and $X_{2}=\mathrm{sgn}(X)\in\{-1,1\}$ (sign of $X$). 
Clearly, if $X$ follows probability distribution $p$, then  $X_1$ has probability distribution $2p(x)$ for $x>0$. Also, $X_2$ is Rademacher distributed (equiprobable $\pm 1$).
 
 One easily computes $H(X_1)=\sum_{x>0} 2p(x)\log \frac{1}{2p(x)} = H(X) - \log 2$ and $H(X_2)=\log 2$ (equiprobable $\pm1$), hence $d(X,X_1)=1-\frac{H(X_1)}{H(X)}=\frac{\log 2}{H(X)}$ and $d(X,X_2)=1-\frac{H(X_2)}{H(X)}=1-\frac{\log 2}{H(X)}$. Therefore,
$d(X,X_1)+d(X,X_2) = 1$: Inequality~\eqref{eq-possible?} is satisfied with equality. 

Of course, in this trivial example, perfect reconstruction is possible since $X=|X|\mathrm{sgn}(X)=X_1X_2$.
Then by Theorem~\ref{thm:necessary-cond}, we deduce that $X_1$ and $X_2$ are independent. This is easily checked directly since  by the symmetry assumption, $\mathbb{P}(X_{1} = x_{1}\mid X_{2} = \pm1) = \mathbb{P}(X_{1} = x_{1})$.
Notice that from this independence, by Theorem~\ref{thm:sufficient}, we find anew that perfect reconstruction of $X$ is possible from $X_1$ and $X_2$.
 

This example can be easily generalized to the case of a ``symmetric'' complex-valued random variable $X$ with modulus $X_1=|X|$ and argument $X_2=\arg(X)$, where $X_1$ is independent of $X_2$ and $X_2$ is uniformly distributed over $M$ possible values. Then $H(X_2)=\log M$, $H(X_1)=H(X)-\log M$ by symmetry, and similar conclusions hold.

Of course, perfect reconstruction $X=X_1X_2$ is always possible even in the case where $X$ is \emph{not} symmetric, in which case $X_1$ and $X_2$ are \emph{not} independent and, therefore, by the alignement condition, $d(X,X_1)+d(X,X_2) =d(X_1,X_2)<1$.

\subsection{Linear Transformation over a Finite Field}

Consider $X$ uniformly distributed over $\mathbb{F}_{q}^{k}$, where $\mathbb{F}_{q}$ be the field with $q$ elements. Suppose $X$ is linearly transformed using some matrix $\mathbf{G}$ to obtain
\begin{equation}
(X_{1},X_{2},...,X_{n})=X\cdot \mathbf{G}
\end{equation}
in row vector notation, where $\mathbf{G}\in \mathbb{F}^{k\times n}_{q}$ has $k$ rows and $n$ columns.
For example, $X$ may represent information symbols to be transmitted over a channel, and $(X_{1},X_{2},...,X_{n})$ would be the associated codeword using an $(n,k)$ linear code over $\mathbb{F}_{q}$ with generator matrix $\mathbf{G}$.

If the $i$th column of $\mathbf{G}$ is not the all-zero vector, then it is easily checked that since $X$ is uniformly distributed over $\mathbb{F}_{q}^{k}$, $X_i$ is likewise uniformly distributed over $\mathbb{F}_{q}$.
Therefore,
\begin{equation}
d(X,X_i)
=1- \frac{H(X_i)}{H(X)} = 1- \frac{\log q}{\log q^k}
= 1- \frac{1}{k}.
\end{equation}
When the $i$th column of $\mathbf{G}$ is all zero, however, $d(X,X_i)=d(X,0)=1$. Summing up,
\begin{equation}
\sum_{i=1}^n  d(X,X_i) = n - \frac{n'}{k}
\end{equation}
where $n'\leq n$ is the number of nonzero columns in $\mathbf{G}$.

By Theorem~\ref{thm:necessary-cond}, if $n-\frac{n'}{k}>n-1$, that is, $n'<k$, then perfect reconstruction is impossible. This is quite natural since in this case, $(X_{1},...,X_{n})$ entails less $q$-ary symbols than the vector $X$, so that it is impossible to reconstruct $X$ from the $n'$ actual symbols in $(X_{1},...,X_{n})$.

In general, if $\mathbf{G}$ has rank $r\leq \min(k,n')$, then since $X$ is uniformly distributed over $\mathbb{F}_{q}^{k}$, the vector $(X_{1},...,X_{n})$ is also uniformly distributed over a subspace of $\mathbb{F}_{q}^{n}$ of dimension $r$. 
Now as we have just seen,  if the $i$th column of $\mathbf{G}$ is not the all-zero vector,  then $X_i$ is uniformly distributed over~$\mathbb{F}_{q}$.
Since uniformly distributed components of a discrete random vector are independent iff the vector is itself uniformly distributed, the only possibility for components $X_{1},...,X_{n}$ to be independent as in Theorem~\ref{thm:sufficient} is that $(X_{1},...,X_{n})$ is  uniformly distributed over $\mathbb{F}_{q}^{n'}$, that is, $r=n'=k$. In this case $\sum_{i=1}^n  d(X,X_i) =n-1$ and by Theorem~\ref{thm:sufficient}, perfect reconstruction is possible. 
Of course, from linear algebra, we known that $X$ can be reconstructed from $(X_{1},...,X_{n})$ as soon as $\mathbf{G}$ has rank $r=k\leq n'$. 


Due to the power of linear algebra, this example may appear quite trivial. 
It would be interesting to generalize it, however, to the case where the vector $(X_{1},...,X_{n})$ 
is obtained by a \emph{nonlinear} transformation, i.e., each $X_i$ are boolean functions over $\mathbb{F}_q$ of the components of vector $X$, e.g. described in algebraic normal form.

\subsection{Integer Prime Factorization}

Consider an integer-valued random variable $X$, uniformly distributed over $\{1,2,\ldots,m\}$, and let $n=\pi(m)$ be the number of primes not exceeding $m$. For every such prime $p$, let $X_p$ be the $p$-adic valuation of $X$, that is, the largest exponent of $p$ such that $p^{X_p}$ divides $X$.
We know by the fundamental theorem of arithmetic that the prime factorization of $X$ always exists and is unique: $X=\prod_p p^{X_p}$, hence $X$ can be reconstructed from the $X_p$'s.

There are $\lfloor\frac{m}{p^{k}}\rfloor$ values of $X$ divisible by $p^k$ and, therefore,
$\lfloor\frac{m}{p^{k}}\rfloor\!\!-\!\!\lfloor\frac{m}{p^{k+1}}\rfloor$ values of $X$  such that $X_p=k$.
Thus, $H(X|X_p=k)=\log \bigl(\lfloor\frac{m}{p^{k}}\rfloor\!\!-\!\!\lfloor\frac{m}{p^{k+1}}\rfloor \bigr)\leq \log \frac{m}{p^{k}}=\log m - k\log p
$, $H(X|X_p)\leq \log m - \mathbb{E}(X_p)\log p$, and
\begin{equation}
\sum_{p\text{ prime }\leq m} d(X,X_p) =
\sum_{p\text{ prime }\leq m} \frac{H(X|X_p)}{H(X)}
\leq \sum_{p\text{ prime }\leq m} \frac{\log m - \mathbb{E}(X_p)\log p}{\log m}
=n - \frac{\log m!}{m\log m}.
\end{equation}
In the latter equality we have used the exact value $\sum_{p} \mathbb{E}(X_p)\log p = \frac{\log m!}{m}$. This can be easily checked from the reconstruction formula itself (!), since 
\begin{equation}
\sum_{p\text{ prime }\leq m} \mathbb{E}(X_p)\log p = 
\mathbb{E} \log \prod_{p\text{ prime }\leq m} p^{X_p} = \mathbb{E} \log X = \frac{\log m!}{m}.
\end{equation}

Since $\log m! \leq m\log m$, the above upper bound is not tight enough to prove inequality~\eqref{eq-possible?} of Theorem~\ref{thm:necessary-cond}.
It is only satisfied asymptotically as $m\to+\infty$ since then $\frac{\log m!}{m\log m}\to 1$.
Likewise, the independence assumption of Theorem~\ref{thm:sufficient} is only true asymptotically: In fact,
since for distinct primes $p_1,\ldots,p_\ell$,
\begin{equation}
\mathbb{P}\bigl(X_{p_1}\geq k_1,\ldots,X_{p_\ell}\geq k_\ell\bigr)   = \frac{1}{m} \biggl\lceil \frac{m}{p_1^{k_1}\cdots p_\ell^{k_\ell}}\biggr\rceil \to \frac{1}{p_1^{k_1}\cdots p_\ell^{k_\ell}}
\end{equation}
it follows that the $X_p$ converge in distribution toward \emph{independent} geometric variables with respective parameters $1-\frac{1}{p}$.

\subsection{Chinese Remainder Theorem}

Consider an integer-valued random variable $X$, uniformly distributed over $\{0,1,\ldots,k-1\}$ where 
$k = \prod_{i=1}^{n}k_{i}$ is the product of $n$ pairwise coprime factors $>1$, and define the following remainders modulo these factors:
\begin{equation} \label{equation-crm}
\begin{cases} 
X_{1} &\equiv X \bmod k_{1} \\
X_{2} &\equiv X\bmod k_{2}\\
&\vdots \\
X_{n} &\equiv X \bmod k_{n}.
\end{cases}
\end{equation}
By the well known \emph{Chinese remainder theorem}, this system of equations has a unique solution in $\{0,1,\ldots,k-1\}$, i.e.. perfect reconstruction of $X$ is possible.


Clearly, since $X$ is uniformly distributed, $X_i$ is likewise uniformly distributed over $\{0,1,\ldots k_i-1\}$ so that $H(X_i)=\log k_i$, $d(X,X_i)=1-\frac{H(X_i)}{H(X)}= 1- \frac{\log k_i}{\log k}$ and
\begin{equation}
 \sum_{i=1}^{n}d(X,X_{i})=   \sum_{i=1}^{n}\bigl(1- \frac{\log k_i}{\log k}\bigr) = n- \frac{\log  \prod_{i=1}^{n} k_i}{\log k} = n-1.
\end{equation}
Thus, inequality~\eqref{eq-possible?} of Theorem~\ref{thm:necessary-cond} is achieved with equality, which proves that $X_1,X_2,\ldots,X_n$ are independent.
Had we proved directly this independence, Theorem \ref{thm:sufficient} would have shown that perfect reconstruction is possible. Thus, a information theoretic proof of the Chinese remainder theorem using this method amounts to proving such an independence. But this can be done quite similarly as the Chinese remainder theorem is classically proved.

With our present method, however, it can be easily seen that perfect reconstruction would \emph{not} be possible if we do not use all components $X_1,X_2,\ldots,X_n$. Indeed, suppose without loss of generality that one tries to reconstruct $X$ only from $X_1,X_2,\ldots,X_{n-1}$. 
Then by the above calculation,
\begin{equation}
 \sum_{i=1}^{n-1}d(X,X_{i})=   \sum_{i=1}^{n-1}\bigl(1- \frac{\log k_i}{\log k}\bigr) = n-1- \frac{\log  \prod_{i=1}^{n-1} k_i}{\log k} = n-2 + \frac{\log k_n}{\log k} >n-2
\end{equation}
which shows by Theorem~\ref{thm:necessary-cond} that perfect reconstruction of $X$ from less than $n$ remainders is impossible.


\subsection{Optimal Sort} \label{optimal-sort}
In this Subsection, we provide a new information theoretic proof of the following 

\begin{Theorem}
Any pairwise comparison-based sorting algorithm has worst-case computational complexity $\geq \log_2 k! = \Omega(k\log_2 k)$ where $k$ is the cardinality of the list to be sorted.
\end{Theorem}

\begin{proof}
Consider a finite, totally ordered list of $k$ elements. It can be seen as a permutation of the uniquely sorted elements, and sorting this list amounts to finding this permutation.
Let $X = (X_{1},X_{2},...,X_{k})$ be a (uniformly chosen) random permutation on $\{1,2,\ldots,k\}$. 

For $i,j \in \{1,\ldots,k\}$ with $i \not= j$, let $X_{i,j}$ be the binary random variable taking the value $1$ if $X_{i} < X_{j}$ and $0$ otherwise. Clearly $X_{i,j}\leq X$ for any $i,j$.



Since there are as many permutations such that $X_{i} < X_{j}$  as such that $X_{i} > X_{j}$,
every $X_{i,j}$ is a Bernoulli ($1/2$) variable (equiprobable bit). Therefore, 
\begin{equation}
d(X,X_{i,j})=1-\frac{\log 2}{\log k! } = 1-\frac{1}{\log_2 k! }.
\end{equation}
Assuming $n$ pairwise comparisons are made to sort the complete list, this gives
\begin{equation}
\sum_{i,j \text{ ($n$ terms)}} d(X,X_{i,j})= n-\frac{n}{\log_2 k! }.
\end{equation}
%
By Theorem \ref{thm:necessary-cond}, it is necessary that this value does not exceed $n-1$, i.e., $n\geq \log_2 k!$ for perfect reconstruction to hold. In other words, the wort-case complexity to achieve the complete sort for \emph{any} possible realization of the initial unsorted list requires at least $\lceil\log_2 k!\rceil$ pairwise comparisons.
\end{proof}

\begin{Remark}
This example illustrates a method to find a lower bound on the worst-case complexity of a problem. The first step is to express the instance of the problem as a random variable $X$. Second, one determines which pieces of information one is allowed to extract from $X$, and models them as ``observed'' random variables $X_{i} \leq X$. Third, for each $i$, we compute the Rajski distance $d(X,X_{i})$. Finally, we use Theorem \ref{thm:necessary-cond} to find a lower bound on the number of ``observed'' variables $X_{i}$ that 	are required to reconstruct $X$. 
We feel that such a method is interesting because it is often harder to find a lower bound on the complexity of a problem than to find an upper bound on it.  
\end{Remark}

\section{Conclusion and Perspectives}

It is an understatement to say that
the ``true'' information theory of 1953 was not as popular as the classical theory of 1948.
John Pierce, a colleague of Shannon, wrote that
``\textsl{apparently the structure was not good enough to lead to anything of great value}''~\cite{Pierce73}.
We find two possible reasons for this pessimism: the fact that the lattice is not Boolean, which does not facilitate the calculations; and the discontinuous nature of the common informtion with respect to the entropy metric.

However, as we have shown in this paper, this lattice structure is quite helpful to understand reconstruction problems. As shown in Section~\ref{application}, the implications of the resolution of perfect reconstruction problems go beyond signal processing, since the concept of information is pervasive in all fields of mathematics and of science. Thus, we believe it is important to deepen this theory, defining \emph{information} \emph{per se}, and to further generalize the reconstruction problems. It would indeed be of great interest to find a simple sufficient condition to reconstruct a variable $X$ from (not necessarily independent) components $X_1,X_2,\ldots,X_n$.  

One may legitimately argue that most examples (except in~Subsection~\ref{S-abssign}) assume uniform distributions, where the entropy is just a logarithmic measure of the alphabet size, and since all considered processings are deterministic, the essence of the present reconstruction problem appears more combinatorial than probabilistic. Indeed, a desirable perspective is to go beyond perfect reconstruction of discrete quantities by considering the possibility of noisy reconstruction of discrete and/or continuous sources of information.

In a perspective closer to computer science, we have used our theorems to find a lower bound on the complexity of the comparison-based sorting problem. It would be interesting to find other problems for which a lower bound on complexity can be found using our technique, especially for decision problems that are not known to be in P.

Finally, as another practical perspective for security problems, one may assume that 
$X$ models all the possible values that can take a secret key in a given cryptographic device, and that an attacker can observe $k$ random values that are deterministically obtained from $X$. Such important problems have been studied e.g., in~\cite{Malacaria15} to evaluate information leakage in the execution of deterministic programs. One may use the theorems of Section~\ref{reconstruction} to find a lower bound on $k$ for the attacker to be able to reconstruct the secret. 

\begin{adjustwidth}{-\extralength}{0cm}

\reftitle{References}


\begin{thebibliography}{99}


\bibitem{Cherry51}
E. C. Cherry, ``A history of the theory of information,'' in \textit{Proc. Inst. Electrical Engineering} (London), vol.~98, no.~55, pp.~383--393, Sept.~1951.

\bibitem{CsiszarKorner11} 
I. Csiszár and J. Körner, \textit{Information Theory. Coding Theorems for DiscreteMemoryless Systems}. Cambridge University Press, 2nd edition, 2011.

\bibitem{Fano52}
R. M. Fano, \textit{Class notes for course 6.574 : Transmission of Information}, MIT, Cambridge, MA, 1952.

\bibitem{Fano01}
\textemdash, ``Interview by Aftab, Cheung, Kim, Thkkar, Yeddanapudi, 6.933 Project History, Massachusetts Institute of Technology,'' Nov. 2001.

\bibitem{GacsKorner73}
P. G\'acs and  J. Körner, ``Common information is far less than mutual information,'' \textit{Problems of Control and Information Theory}, vol.~2, no.~2, pp.~149--162, Jan.~1973.

\bibitem{Horibe73}
Y. Horibe, ``A note on entropy metrics,'' \textit{Information and Control}, vol.~22, no.~4, pp.~403--403, May 1973.


\bibitem{Jaccard01}
P. Jaccard, ``Distribution de la flore alpine dans le bassin des Dranses et dans quelques régions voisines'', \textit{Bulletin de la Société Vaudoise des Sciences Naturelles}, vol.~37, no.~140, pp.~241--272, 1901. 

\bibitem{Malacaria15}
P. Malacaria, ``Algebraic foundations for quantitative information flow,'' \textit{Mathematical Structures in Computer Science}, vol.~25, no.~2, pp.~404--428, Feb. 2015.


\bibitem{Nakamura70}
Y. Nakamura. “Entropy and semivaluations on semilattices,” \textit{Kodai Math. Seminar Report}, vol.~22, no.~4, pp. 443--468, 1970.

\bibitem{Pierce73}
J.R. Pierce, ``The early days of information theory,'' \textit{IEEE Transactions on Information Theory}, vol.~19, no.~1, pp.~3--8, Jan. 1973.

\bibitem{Rajski61}
C. Rajski, ``A metric space of discrete probability distributions,'' \textit{Information and Control}, vol.~4, no.~4, pp.~371--377, Dec. 1961.

\bibitem{Rioul07}
O. Rioul, \textit{Théorie de l'information et du codage}, Hermes Science - Lavoisier: London, 2007.

\bibitem{Gretsi22}
O. Rioul, J. Béguinot, V. Rabiet and A. Souloumiac, ``La véritable (et méconnue) théorie de l'information de Shannon'', in Proc. \textit{28e Colloque GRETSI 2022}, Nancy, France, 6--9 Sept. 2022.

\bibitem{Shannon48}
C. E. Shannon, ``A mathematical theory of communication,'' \textit{Bell Syst. Tech. J.}, vol. 27, no. 3 \& 4, pp. 379--423 \& 623--656, July \& Oct. 1948.

\bibitem{Shannon50}
\textemdash, ``Some topics on information theory,'' in \textit{Proc. Int. Congress Math.}, AMS, Ed., vol. \cRM{2}, Aug. 30--Sept. 6, 1950, pp. 262--263.

\bibitem{Shannon53}
\textemdash, ``The lattice theory of information,'' in \textit{Transactions of the IRE Professional Group on Information Theory}, vol. 1, no. 1, pp. 105--107, Feb. 1953.

\bibitem{Shannon56}
\textemdash, ``The bandwagon (editorial),'' \textit{IRE Transactions on Information Theory}, vol. 2, no. 1, p.~3, Mar. 1956.

\bibitem{Wyner75}
A. D. Wyner, ``The common information of two dependent random variables,'' \textit{IEEE Transactions on Information Theory}, vol. 21, no. 2, pp. 163--179, Mar. 1975.

\bibitem{Yeung08}
R. W. Yeung, \textit{Information Theory and Network Coding}. Springer, 2008.

 


\end{thebibliography}

\PublishersNote{}
\end{adjustwidth}

\end{document}